# A Review on the Properties and Nature of Light

**Prachurjo Dutta Roy**[1]

***Abstract*** – The following paper is a detailed review on the properties and nature in relation to the phenomena known as 'light'. The paper tries to chronologically detail the development of theories and related experiments from the philosophical Greco-Roman and Indian schools of thought to the modern quantum theories.

## 1. INTRODUCTION

What is Light? Light is the everyday phenomena that helps us to visualise and see things. For a long while, and even now, philosophers and scientists alike have been struggling to understand the nature of this illuminating entity.

The study of light came about as a result of trying to understand the idea of vision and colours as discussed by Crone in [1]. Crone outlines the Hellenistic ideals of 'visual rays' or 'optic rays' proposed by Pythagoras, which he thinks to be some form of fine matter[2] emerging in a straight line from the eye and reaching the object. The Greek craze about geometry and straight lines aided this idea.

A contemporary of Pythagoras, Empedocles motivated by the four elements of antiquity, believed that Aphrodite made the human eye out of the four elements and the fire inside the eye allowed sight. This suggested equal vision in both day and night, so Empedocles postulated that interaction occurred between the fire inside and certain objects (sources) like the sun [2]. Empedocles suggested that light must have a finite speed, as it was something in motion and as a result would take time to travel through space.

Around the same time in India, two philosophical schools, the *Samkhya* and the *Vaisheshika* led the ideas on optics and vision. The *Samkhya* school suggests light as a fundamental element along with the four elements of antiquity. The *Vaisheshika* school, however, takes an atomistic route, suggesting that light is actually formed due to the motion of high-velocity fire (*tejas*) atoms. There is further differentiation on the particles of light, depending on the speed and arrangement of the fire atoms [2].

Atomism and theories of vision have also gone hand to hand in Hellenistic philosophies. It is here that we come across the ideas of Intromission and Extramission[3]. Intromission refers to the concept that vision takes place from the object to the eyes and Extramission refers to when vision takes place from the eyes to the object. For a long time, the predominant idea was of Extramission, based on which Heron suggested that the speed of light must be infinite, as one could see stars at the instant of opening one's eye, but Democritus' theory of atoms, gave the idea that small infinitesimal particles (*atomons*) travelled from the world into the eye. It is here that we see references to the fact that objects form images in the eye via reflection [1] which would suggest a finite speed like in Empedocles' model.

Crone then goes on to discuss Aristotle and his definition of 'change'[4] in [1, p. 10], "…Aristotle postulates a conducting medium, the air. But the stars are seen through a different medium, the aether. This is the fifth, divine element, *quinta essentia*. The medium in the eye itself is the eye fluid. In this transparent medium light actualises the potential of colours to affect the eye. A distinctive feature of this train of thought is that actualisation is being considered and not movement in space. The actualisation occurs twice: in the first place the transparent medium is actualised - that is light." Light was further defined then as an *accidens*, a state of something else.

Due to Alexander's conquests, the Greco-Roman ideas on optics spread to Persia and surrounding areas and had a profound impact during the Golden Age of Islam.

Ibn Sahl, one such scholar, experimented with curved lenses and mirrors which led him to a formula similar to the Snell's Law [3],

---

1 Delhi Public School, Ruby Park, Kolkata

2 Not to be confused with atoms or matter particulates. Matter was thought to be continuous in most Greek schools of philosophy

3 The concepts were known but not well-developed until the ideas of Atomism came into being.

4 Change is said to be related to three stages: potential, development and actualisation.

$$\mu_1 \sin\theta_1 = \mu_2 \sin\theta_2 \quad (1)$$

Where $\mu_1$ and $\mu_2$ are the refractive indices of the media 1 and 2 and the $\theta_1$ and $\theta_2$ are the incident angles for the media 1 and 2. Ibn Sahl had also managed to form lenses using the law to focus sun rays and burn objects.

Another Middle Eastern scholar and Ibn Sahl's successor, Alhazen[5], in his *Kitab al-Manazir*[6] does a more pinpointed study on the behaviour of light and the science behind vision. The Intromission theory is concretised by the fact that eyes can be damaged by looking directly at bright lights, such as the sun [4]. Alhazen postulates that every object in the universe gives off some light from every point on its surface, which travels to the eye and allows vision of that object. This basically suggests that there are infinite rays emitted from the object that allows the object to be viewed. The Pythagorean idea of light travelling in straight lines is also seen to be entertained in his theories.

Alhazen uses the *camera obscura* (pinhole camera) to propound that infinite light rays come together at a single point and form images without any colour distortion. He uses this idea to define how light enters the eye from the outside world, forms an 'image' and allows us to see [1]. To distinguish the directions of the infinite rays, Alhazen suggests that light that enters the eye is not refracted[7] [1] [5].

Existence of primary and secondary light is also claimed in the *Kitab al-Manazir*, with the first being more intense than the latter [4] [5]. It is suggested that primary light (essential form) comes from self-luminous bodies and secondary light is the light that comes from accidental objects [5].

Another property discussed is that of transparency. Transparent objects are those that allow light to transmit through itself (e.g., Air, Water). Opacity is defined to have various degrees, with opaque bodies themselves turning into secondary luminous objects which give off secondary light. Light is said to refract when passing through partially transparent objects and reflect when striking a smooth surface (e.g., Mirrors, Polished Metals). Selin suggests that Alhazen is the first to arrive at the conclusion that during reflection, the incident ray $i$, reflected ray $r$ and the normal all fall on the same plane [6].

Alhazen also draws similarities between colour and light, suggesting that both are distinct properties of a form and travels from every point on an object in straight lines.

By now, concepts surrounding to light, refraction, reflection, vision and colour had started to be moulded by scientific thinkers and philosophers alike.

## 2. RENAISSANCE & THE SCIENTIFIC REVOLUTION

Many Western scholars had attempted to bring the ideas of Antiquity and Islamic ideas into one coherent unified working theory. The most notable of such scholars is Roger Bacon. Bacon uses Alhazen's theory and starts to build on it. The most daunting of tasks of Bacon was to unify the differing theories of Empedocles, Alhazen, Aristotle, *et al.*, under one roof. He uses the Intromission theory as his basis, but doesn't rule out the existence of rays emerging from the eyes. He suggests such rays were not responsible for vision but prepared the media required by the incoming rays from the object. He also agrees to the Neoplatonic idea that the universe was a vast network of forces and both light and colour were particular manifestations of the said universal force [5]. He also suggests that the speed of light was finite, by using the arguments of Alhazen and Aristotle.

Alongside Bacon, two more scholars, Pecham and Witelo propagated Alhazen's theories into the western world which made Alhazen's ideas the go-to for any epistemological discussions pertaining to light [5]. Witelo considered that lightspeed could be infinite in vacuum, but slowed down in denser objects.

Leonardo da Vinci is the first to compare the eye to the *camera obscura*. As a result of which, he is able to disprove Galen's theory that the lens is not what enables vision, but an optical instrument. The lens ensures that the light rays form the image (upright) on the required spot of the eye (optic nerve). A notable idea seen in such theories is the avoidance of inverted images being formed inside the eyes [1].

After Vinci, a few others tried to add to the now growing field of optics. An important figure in this interval was a Sicilian priest, Francesco Maurolyco. Maurolyco was the first to describe the action of prisms and the phenomenon of chromatic aberration [1]. Another figure, Giambattista Della Porta works with prisms and Crone provides a diagram by Della

---

5 Also known as Ibn al-Haytham.

6 Read the translation given in the references section [4] for a more thorough understanding.

7 Refer to the diagram provided by Crone [1, p. 23] or Lindberg [5, p. 312]

Porta [1, p. 43], showing refraction of sunlight via a semi-transparent glass. The colour blue travels the longest, while yellow travels the shortest.

Following Vinci, the next major name is of Johannes Kepler. Kepler studies the images formed by the moon using a *camera obscura* and comes to the conclusion that the eye must depend on a different mechanism, i.e. refraction [1]. This disproves the fact that there is no single ray that is responsible for vision and all rays converge at a single point.

Studies on refraction and telescopes led Kepler to a relation between $i$ and $r$,

$$i = nr \quad (2)$$

This empirical observation was valid for angles ≤ 30° and it was seen that for light travelling from air to glass, the value of the constant $n$ was equal to 1.5 [7]. Eq. 2 is also a special case of Eq. 1 as for small angles, $\sin\theta \cong \theta$.

Kepler provides another breakthrough, albeit with hesitance (a recurring theme that occurs as history progresses), as Crone [1, p. 45] suggests, "His experiments with liquid-filled globes confirm that good convergence of light rays and adequate image formation are possible if the rays enter through a comparatively small opening […] The eye possesses such a diaphragm: the iris. The eye also has refracting components such as the lens, which can bring the rays together again to one point. […] The obvious place […] was the retina. Kepler relates how he tortured himself until he was nearly out of his mind, in an attempt to construct (in Leonardo's manner) double crossing of the rays."

His futile attempts finally led him to a fact that the likes of Alhazen and Vinci could not accept – images formed in the retina were inverted. Kepler also took the side of Heron and says the speed of light is infinite as empty space does not act as an obstacle to it.

Kepler also finds out the inverse-square law of light,

$$I \propto \frac{1}{r^2} \quad (3)$$

Here $I$ is the intensity of light and $r$ is the distance travelled by it.

By the time of Kepler, the small sparks of scientific revolution had begun to burn brightly, and the man who achieved great heights due to this was René Descartes. Cartesian ideas now began to change how thought worked and the man who had an opinion on everything also had opinions on light and vision.

The Cartesian theory of light started with the *Dioptrique*, the first of the three treatises by Descartes in the *Discours*. Descartes takes the following ideas into account:

1. Light itself was nothing but a mechanical property of the luminous object and of the transmitting medium [8]. This medium, the *plenum* is a reintroduced form of the Aristotelian *quinta essentia*, based on the *horror vacui*[8] concept.
2. Light is a certain movement or action[9], very rapid and very lively (almost like a 'pulse' or 'pressure'), which passes toward our eyes through the medium of the air and other transparent bodies [9] in a rectilinear manner and also extends around in all directions about bodies one calls "luminous" [10].
3. The propagation of light was instantaneous, i.e., that lightspeed was infinite [8] [10] which he argues from the fact that if the speed of light were to be finite, the Sun, Earth, and Moon would be noticeably out of alignment during a lunar eclipse [8] [11]. He compared this to the fact that any pressure applied on a stick at one end translates to the other end instantaneously [8].
4. The rays of light can be divergent or convergent [10].
5. The intensity of such rays can be strengthened or diminished [10].
6. The law of reflection[10] is given by an analogue of a tennis ball hitting a surface [8] [9] [10].
7. The law of refraction (which is the same as Eq. 1) is given by an analogue of a tennis ball puncturing thin sheets of cloth [9] [10], which he reaches via three assumptions based on his idea of momentum conservation [7]:
    a. The speed of light was faster in a denser media.
    b. For the same media, the velocities $v_1$ (pre-refraction) and $v_2$ (post

---

[8] The concept basically states that nature abhors a vacuum and therefore, voids cannot exist.

[9] Sabra [8, p. 50] states, "For, according to him, light does not consist in an actual motion, but rather in a 'tendency to motion' which is transmitted to the eye through a medium."

[10] The relation $\angle i = \angle r$ was known since Euclidean times, but many interpretations of the law were given across history.

refraction) have a constant ratio, independent of $i$.

c. The velocity component parallel to the refracting surface remains unchanged.

The mathematical formulation of the reflection law was explained by comparing it to an elastic collision.

$$v_i = v_r \tag{4}$$

Descartes assumes that the incident velocity $v_i$ and the reflected velocity $v_r$ are conserved.

To accommodate the velocity components and angles, he then assumes the following,

$$v_i \sin i = v_r \sin r \tag{5}$$

From Eq. 4 and Eq. 5 we get,

$$\sin i = \sin r$$

or,

$$i = r \tag{6}$$

It should be noted that Descartes does not postulate about the perpendicular component of the incident velocity in his proof [8].

A mathematical formulation can also be provided to assumption 7b,

$$\frac{v_r}{v_i} = n \tag{7}$$

The constant ratio $n$ is basically the refractive index of medium 1 w.r.t. medium 2 ($\mu_{12}$). Although it must be noted that the nature of the constant was not known during the time of Descartes' proof.

Descartes presents Eq. 7 as,

$$v_r = nv_i$$

Here $n$ is obviously $\mu_{12}$, which Descartes considers in his proof to be an arbitrary constant. He reapplies Eq. 5 and it is given that $v_1 = v_r$ and $v_2 = v_i$. So, from Eqs. 5 and 7 we get,

$$\sin i = n \sin r \tag{8}$$

Plugging $n = \mu_{21}$, basically gives us Eq. 1, the Snell's Law [8].

A few years before, Willebrord Snell, an astronomer and mathematician, via empirical experimentation, arrives at the Snell's Law (Eq. 1), although in an inconvenient form [7],

$$\frac{\csc \theta_1}{\csc \theta_2} = \mu_{12} = \frac{1}{\mu_{21}} = constant \tag{9}$$

However, Snell never published his findings [7]. Here starts a controversial period between that of Descartes and Fermat that takes the study of light to the next stage.

It is in the Cartesian Law of Refraction (albeit correct in the mathematical sense) that Descartes' *méthode* fails to work. There is a clear contradiction between the tennis-ball analogy and assumption 7a. The inaccuracies of these theoretical assumptions[11] brought upon attacks and questions by his contemporaries like Fermat and Rømer [7].

Pierre de Fermat first criticised the derivation of the Cartesian law of reflection. Sabra comments, "Fermat could not accept Descartes' argument for establishing the law of reflection as a 'legitimate proof and demonstration'. To his mind there was something in this argument that was logically wrong; and his attack was directed against the method of resolving motions that was adopted by Descartes in his treatment of both reflection and refraction. For Fermat believed that, in resolving the motion of the incident ball into those particular components, the one perpendicular and the other parallel to the reflecting surface, Descartes had chosen but one out of an infinite number of reference systems which were all valid. But since he had chosen the one which could help him to arrive at the desired, and already known result, his procedure was in Fermat's view completely arbitrary." [8, p. 87]

Fermat then goes after the law of refraction by stating the pressure-like stick analogy and the corpuscular moving-ball analogy were contradictory. Fermat himself states, "…it seems that there is a particular disproportionality in that the motion of a ball is more or less violent, according as it is pushed with different forces, whereas light penetrates the diaphanous bodies in an instant and seems to have nothing successive in it." [8, p. 112]

As Sabra continues to elucidate, "But how are we to account for this assumption[12] in [*sic*] Descartes? […] the apparently *ad hoc* character of this assumption which has led several writers to suspect Descartes of having artificially constructed his deductive proof of the sine law after he had obtained this law by different means. If these means were valid, why would he have suppressed them? Ought we therefore to conclude (with Leibniz and Mach) that he must have derived the law in the first place from Snell's manuscript?" [8, p. 115]

Fermat continues years of correspondence with Cartesians such as Clerselier refuting and trying to disprove Descartes' theory and build his own law of refraction.

[11] Descartes' work on colours and rainbows, however, was heavily based on experimentation. I suggest reading Sabra [8, pp. 62-68] for a more thorough understanding.

[12] The assumption in question is of the re-usage of Eq. 5.

Fermat posits that rays of light must take the shortest path to travel. The concept of nature taking the most simplistic of solutions is a common theme across epistemology, such as the Occam's Razor. There are two problems that are encountered when dealing with shortest paths, 1) in reflection at concave surfaces, the light sometimes followed a longer path and 2) the principle was not applicable to the behaviour of light in refraction: obviously, if the light always followed the shortest path, it should always go in a straight line from one medium into another [8].

Fermat tries to solve this by suggesting that instead of the 'shortest path', light rays take the 'easiest course' and as a result came to be known as the *principle of least time*. Fermat assumed that the principle would lead him to a new law, different to that of Descartes.

Using his minima-maxima method, Fermat arrived via a rigorous geometric and algebraic proof[13] back to the Cartesian ratio of sines.

$$\frac{\sin i}{\sin r} = K \qquad (10)$$

This result greatly surprises Fermat and he supposedly 'abandons' the battlefield against the Cartesians on his arrival of the result [8]. He then uses the Cartesian relation,

$$\frac{R_r}{R_i} = \frac{v_r}{v_i} = K, \qquad K > 1 \qquad (11)$$

$R_i$ and $R_r$ are the 'resistances' of the incident and refractive mediums respectively. Combining Eqs. 10 and 11, we get,

$$\frac{\sin i}{\sin r} = \frac{v_r}{v_i} \qquad (12)$$

Which is just Eq. 5. This however, presented several paradoxes as Fermat's principle was based on the fact that,

$$R \propto \frac{1}{v} \qquad (13)$$

So, it was impossible that the Cartesian idea of light speeding in denser mediums and his own idea of light slowing down in denser mediums would yield the same result.

As Sabra states, "What he overlooked was, strangely enough, that his principle *does not*, strictly speaking, yield the same law which is obtained in Descartes' proof, though both laws assert the constancy of the ratio of the Sines." [8, p. 149]

What Fermat actually should have obtained via the implications of Eq. 13 was,

$$\frac{\sin i}{\sin r} = \frac{v_i}{v_r} \qquad (14)$$

Fermat could not accept the fact he and Descartes should arrive at the same conclusion via two entirely opposite assumptions. I, for one, sympathise with Fermat on missing Eq. 14 after a rigorous derivation for finding a law of refraction that was true to nature, which he indeed does.

It must be noted that the proof never really applied a quantity of 'time' and is highly confusing in nature. Hence, I shall show a differential proof of Snell's Law via Fermat's principle.

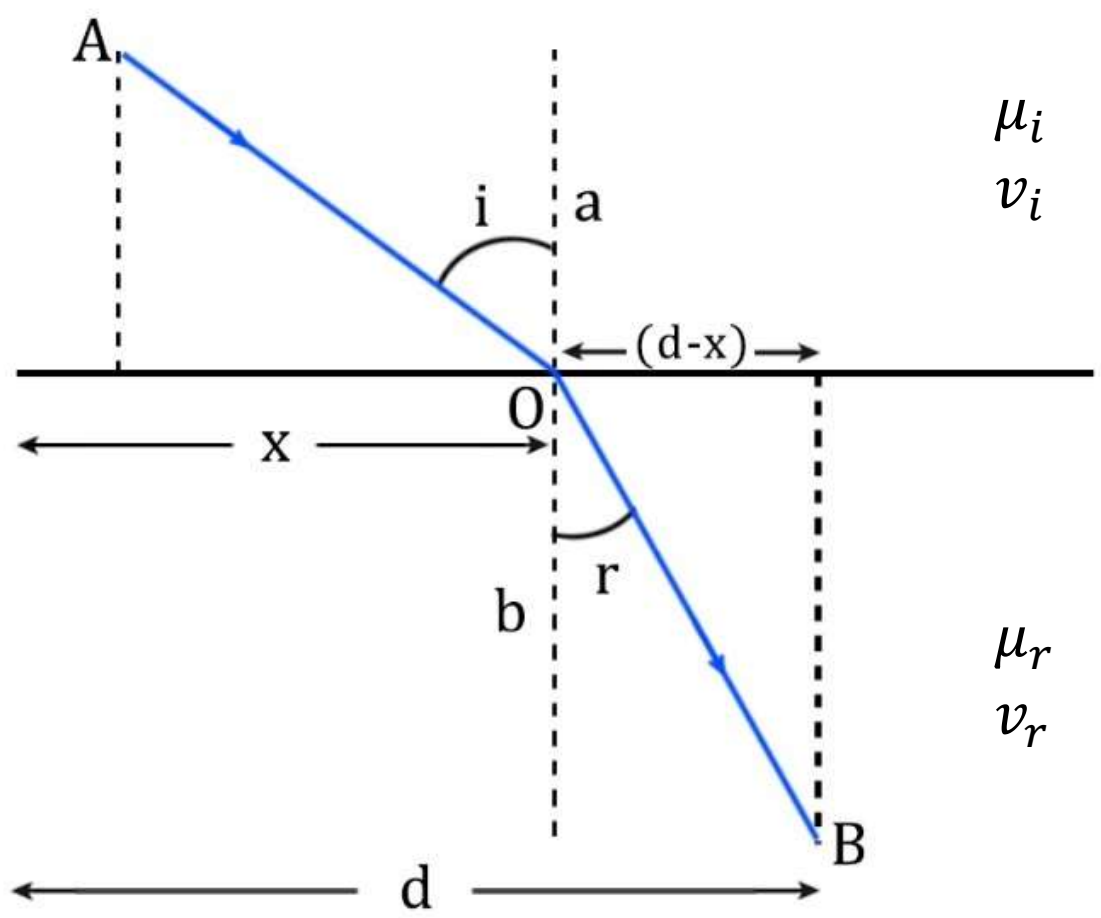

*Figure 1: A light ray **AO** passing through rarer medium **i** refracts via path **OB** in denser medium **r**.*

From the principle, we know, that time taken by a light is minimum,

$$t = T_m \qquad (15)$$

And as per Figure 1, we consider the total time travelled $T$ to be equal to time taken in medium $\mu_i$ and medium $\mu_r$,

$$T = T_i + T_r \qquad (16)$$

From Figure 1, we see $T_i$ is equal to the distance travelled by the velocity $v_i$ in that medium,

$$T_i = \frac{AO}{v_i} = \frac{\sqrt{x^2 + a^2}}{v_i} \qquad (17)$$

And, similarly for $T_r$,

$$T_r = \frac{OB}{v_r} = \frac{\sqrt{(d-x)^2 + b^2}}{v_r} \qquad (18)$$

Therefore, $T$ comes as,

$$T = \frac{\sqrt{x^2 + a^2}}{v_i} + \frac{\sqrt{(d-x)^2 + b^2}}{v_r} \qquad (19)$$

[13] The proof is provided by Sabra [8, pp. 145-147].

We minimise $T$, via the Lagrange multiplier form,

$$T_m = T - \lambda(a + b - D) \qquad (20)$$

Here, $\lambda$ is the Lagrange multiplier and the term $a + b - D$ represents the equality constraint, with $D$ being the total distance travelled. We now partially differentiate $T_m$ with $a$ for medium $i$ via Eq. 19,

$$\frac{\partial T_m}{\partial a} = \frac{\partial}{\partial a}T - \frac{\partial}{\partial a}\lambda(a + b - D)$$
$$= \frac{a}{v_i\sqrt{x^2 + a^2}} - \lambda = 0 \qquad (21)$$

Similarly, partially differentiating with $b$ for medium $r$,

$$\frac{\partial T_m}{\partial b} = \frac{\partial}{\partial b}T - \frac{\partial}{\partial b}\lambda(a + b - D)$$
$$= \frac{b}{v_r\sqrt{(d - x)^2 + b^2}} - \lambda = 0 \qquad (22)$$

We now equate Eq. 21 and 22,

$$\lambda = \frac{b}{v_r\sqrt{(d - x)^2 + b^2}} = \frac{a}{v_i\sqrt{x^2 + a^2}} \qquad (23)$$

And from Fig. 1,

$$\sin i = \frac{a}{\sqrt{x^2 + a^2}} \qquad (24a)$$

$$\sin r = \frac{b}{\sqrt{(d - x)^2 + b^2}} \qquad (24b)$$

Combining Eqs. 24a and 24b with Eq. 23,

$$\frac{\sin r}{v_r} = \frac{\sin i}{v_i} \qquad (25)$$

And a simple rearrangement gives us Eq. 14, a variant of Snell's Law.

Another big result of Fermat's idea was the re-emergence of the notion of a definite speed of light. Several attempts were made by now to find a definitive number describing this speed independently of Fermat.

The first proposed idea is attributed to Beeckman[14], where the time delay $\Delta t$ in seeing a cannon flash one mile away in a mirror by a person with their back to the cannon is used to measure the speed of light [12].

$$\frac{1\, mile}{\Delta t} = v_{light} \qquad (26)$$

A similar experiment was also devised by Galileo Galilei, where two observers, A and B, with lanterns are at a considerable distance from each other, the time which observer A measures from uncovering his lantern until he catches sight of the light of B's lantern is the time which it would take the light to travel from A to B and from B back to A. The result of this experiment was inconclusive [13].

The mathematical formulation involved the time instants $t_A$ and $t_B$ for each observer. These were given by [13],

$$t_A = 2\frac{s_A}{v_{light}} + \tau_A \qquad (27)$$

and,

$$t_B = 2\frac{s_B}{v_{light}} + \tau_B \qquad (28)$$

$\tau_A$ and $\tau_B$ were experimental time delays brought on by human error. If the time delays were same, i.e., $\tau_A = \tau_B$, and we subtract Eq. 28 from Eq. 29, we get,

$$t_B - t_A = 2\frac{s_B}{v_{light}} - 2\frac{s_A}{v_{light}} \qquad (29)$$

or,

$$v_{light} = 2\frac{s_B - s_A}{t_B - t_A} \qquad (30)$$

If we take $v_{lig} \;\; = c_{air}$ and $s_B - s_A$ = 1 mile as taken by the *Academia del Cimento* of Florence in their experiments and place it in Eq. 29, then we would get a time delay of about 1.07394 × $10^{-5}$ s or 10.74 μs, which would make it seem almost instantaneous.

Thus, it was established that the distance between the two observations must be increased, and it is through the Galilean telescopes where the first notable calculation by Ole Rømer is made.

Rømer detected certain variations in the apparent period of revolution of Jupiter's moon, Io. He was convinced that the orbital period of Io was not changing and therefore concluded from his observations and theoretical considerations that the variations were caused by the fact that light has a finite velocity [14].

Rømer's observations led him to predict that the emergence of Io on 16 November 1676[15] would be observed about ten minutes later than would have been calculated by the then-prevalent method. In another paper, he showed that the time taken by light to cross the diameter of Earth's revolutionary orbit was about 22 minutes [15].

This suggests that the time taken by light to reach Earth from the Sun was roughly 11 minutes and gives us a value of,

---

[14] Beeckman and Descartes clashed on the nature of the speed of light to the point that Descartes stated that if Beeckman was correct then he knew nothing of philosophy [8] [12].

[15] There was no actual emergence on the 16th, but one happened a week before with a 9.5m delay [14].

$$v_{ligh} \approx \frac{1\ \text{AU}}{660s} \cong 226{,}663\ kms^{-1} \qquad (31)$$

It must be noted however, that Rømer never actually suggested a numerical value to lightspeed but his tabular data suggested that the speed had a lower limit of $1.4 \times 10^8$ m/s [16]. However, reception of this was slow, with the scientific community divided on the wake of this revelation, just like they were on the fundamental theory of light. Two theories formed in an attempt to explain so – the Corpuscular Theory and the Wave Theory – both seemingly contrasting to one another.

## 3. CORPUSCULAR THEORY

The atomic idea of light as seen in Section 1, began with Empedocles and continued with Atomists like Democritus. During the Scientific Revolution, works of atomists took precedence over Aristotelian physics with notable scientists like Boyle, Newton arguing for a mechanical and fundamental model of matter and non-matter.

The first notable proponent of this theory is Pierre Gassendi, who suggested that everything in the universe was made up of 'corpuscles'[16] and similarly, light is a property carried by particular atoms (*atomi lucificae*) that are identical with heat atoms. These tend to travel at greater than average velocity because they generally have fewer obstacles in their paths than most atoms [17]. This was different from Descartes' pressure-model of light but in ways similar to his ball-model of light [8].

Inspired by Gassendi's works at a very young age, Isaac Newton develops his own hypothesis, disregarding the *plenum* and suggest the following [18] in his book *Opticks*[17]:

1. The phenomenon of reflection and refraction was geometric in nature, and hence, could only be explained by a corpuscular nature of light.
2. All luminous objects emit these 'light corpuscles' into the medium surrounding them.
3. These corpuscles have no mass and are perfectly rigid and elastic.
4. They travel from their source to endpoint with a finite speed via spaces between other corpuscles. Laplace had suggested that there may exist bodies which do not allow to light to escape[18].
5. For refraction, the speed of light is different in different transparent materials.
6. The speed of these corpuscles was much larger in a denser medium than in a less dense one due to higher gravitational pull.

Alongside this, Newton fully explained two phenomena, dispersion of white light and polarisation.

Newton's experiments with sunlight and colours is one of his most famous. Newton used a *camera obscura*, a prism and a dark chamber. He focused a single beam of light on to a prism and changed its angle. After a little movement, a band of colours with red at the top – suggesting least refraction – and violet at the bottom – most refraction – appeared [18]. Newton hypothesised that sunlight or white light was *polychromatic*, i.e., composed of multiple colours and each of these colours were differently refrangible. Hence, a relation was found between the red and violet ends of the light spectrum,

$$\mu_{violet} > \mu_{red} \qquad (32)$$

Newton had tried to determine the velocities of colours within white light. And via Eq. 32 and the fact,

$$v_i \mu_i = v_r \mu_r = c \qquad (33)$$

We get,

$$v_{red} > v_{violet} \qquad (34)$$

Newton did hold a belief that since red refracts the least, it should move faster the most. This was his guiding idea in the earliest stages to explain dispersion via his polychromatic theory.

Newton then happened to realize that there was a very simple observation that might test whether different colours moved with different velocities. Observations of colours of Jupiter's satellites could be used to test the velocity model. Conversations with Flamsteed did not prove to be conclusive of this, however and Newton abandons a velocity-based model in the public forum, while basing his work on it [19].

Pedersen suggests (citing Bechler and Newton) [19],

"NEWTON abandoned the velocity model and turned to a mass model of dispersion. He was never again to use in public velocity in connection with the motion of light corpuscles. […] NEWTON was to outline a general model of light, he would suppose that whatever light be "it consists of successive rayes differing from one another in contingent circumstances, as bignes, forme or vigour". NEWTON would persistently neglect to mention velocity in any general exposition of a light model."

---

[16] Corpuscles are different from atoms in the way that they are divisible into more fundamental units in contrast to an atom's indivisible nature.

[17] The book was published in 1704 after Robert Hooke's death, with whom Newton was having a public feud over the nature of light.

[18] A precursor to the ideas of black hole. Laplace, however, withdrew this claim when Wave Optics started to take over.

Newton then adopts a mechanistic mass-based model of light dispersion, in-line with Descartes and Boyle. He treats the corpuscles as projectiles as both light corpuscles and projectiles generate the Ratio of Sines when they enter a field of force and, after having traversed it, move out [20].

Pedersen shows that by consequence of the Cartesian Ratio, Newton had implicitly suggested that for general cases, the ratio provided a function of the incident velocity [19],

$$\frac{\sin i}{\sin r} = f(v_i) \tag{35}$$

Bechler provides a figure showing the motion of a light corpuscle during refraction,

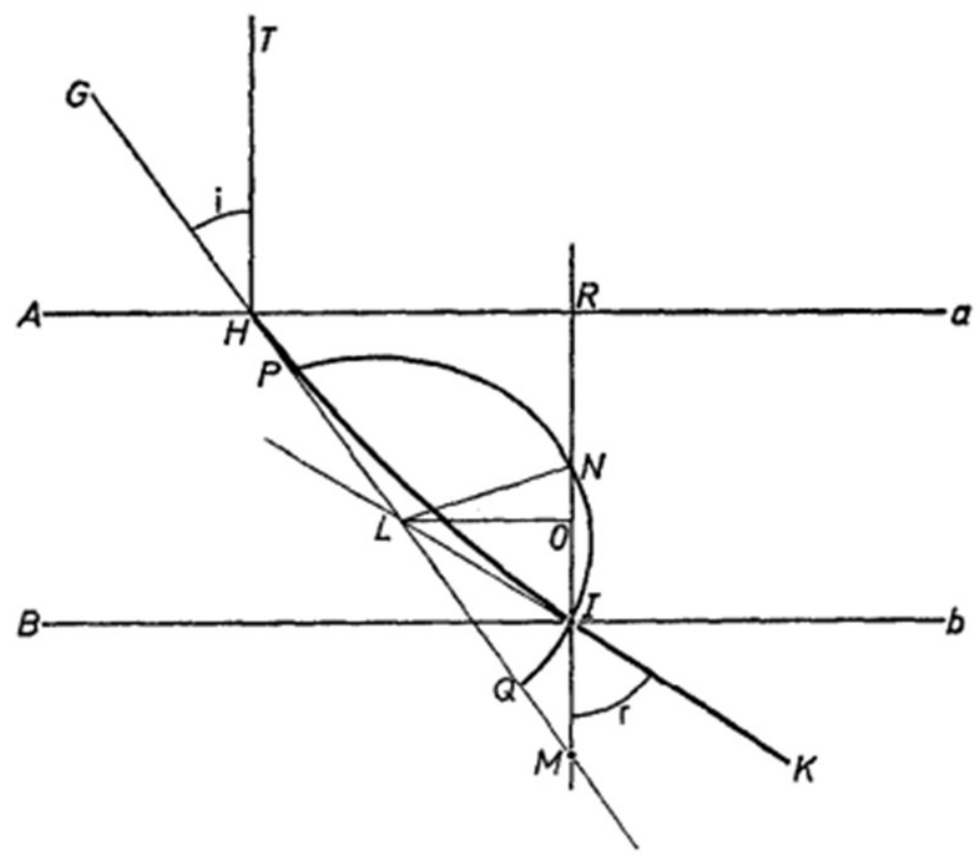

*Figure 2: Light refracting via parabolic path **HI** in a field with boundaries **Aa** and **Bb*** and range ***IR**. [20, p. 15]*

The corpuscle is found to follow the trajectory of a parabola (Figure 2) and the equation for the latus rectum $p$ of the same parabola is represented by,

$$\frac{HM^2}{MI} = 4a = p \tag{36}$$

Via a few more 'geometrical data' (1-3) in [20, p. 15], we arrive at the equation,

$$\frac{\sin i}{\sin r} = \sqrt{1 - \frac{4MN}{p}} \tag{37}$$

From mechanics,

$$HM = v_i t \tag{38}$$

$$MI = \frac{1}{2} g t^2 \tag{39}$$

Placing Eqs. 38 and 39 in Eq. 37,

$$\frac{2v_i^2}{g} = p \tag{40}$$

And we plug that and geometrical data 2, i.e., $MN = IR$ into Eq. 37,

$$\frac{\sin i}{\sin r} = \sqrt{1 - \frac{2g \cdot IR}{v_i^2}} \tag{41}$$

Now, $IR$ is the range of the force field and is invariant of the material. Bechler suggests that $g$ is a function of the distance $r$ but remains constant as $IR$ is constant [20].

$$2\int_0^{IR} g(r)\,dr = 2g \cdot IR = k \tag{42}$$

Combining Eqs. 41 and 42,

$$\frac{\sin i}{\sin r} = \sqrt{1 - \frac{k}{v_i^2}} \tag{43}$$

Eq. 43 gives us the required function in Eq. 35.

It must be noted that in both [19] and later on in [20], an alternate formula is given,

$$\frac{\sin i}{\sin r} = \sqrt{1 + \frac{k}{v_i^2}} \tag{44}$$

For polarisation, Newton stated that the corpuscles were not spherical, but had sides, leading to change in intensities of light on rotation of a polaroid.

Newton's opposition to wave theory was based on the fact that light couldn't bend around obstacles. However, to explain diffraction, Newton suggested the formation of a localised wave in the *aether* medium.

In 1815, Jean-Baptiste Biot demonstrated that "polarised light", when passing through an organic substance, could be rotated clockwise or counter-clockwise, dependent upon the optical axis of the material. What he basically showed was the optical activity in conformers (DL pairs).

The corpuscular theory cemented its authority into to the 19th century due to Newton's influence. It completely broke down with an experiment by Léon Foucault in 1850, where he tried to ascertain the lightspeeds in water and air. He used a beam of light that he passed through a rotating mirror into two tubes – air and water – and studied the images – one clear and one green – formed. He thus concluded that light took longer durations to travel via water and point 6 was proved wrong [21].

## 4. WAVE THEORY

The Wave Theory existed in fragments before the time of Descartes, with he himself suggesting pulse-like or pressure-like behaviour of light.

Francesco Grimaldi is the first to notice[19] in his book, *De Lumine*, an anomalous shadow (of regions of red and blue) [1] cast by a small opaque body, in a sun ray, from a very small aperture. He thus concludes that light not only travels via propagation, reflection and refraction, but also diffraction [22] [23].

Grimaldi disapproves of the corpuscular nature of light and pursues a wave theory of light but in the end, accepts the Aristotelian *accidens* concept. He also posits that the colours formed suggest that colour is a definitive property or an alternate form of light [1].

Robert Hooke, in his book on microbiology and microscopes, *Micrographia*, constructs the first notable theory suggesting wavelike nature of light and suggests the following:

1. There are three properties pertaining to the motion of light [24, pp. 55-56].
    a. It is "exceedingly quick" like that of the motions of fermentation and putrefaction. In all extremely hot shining bodies, there is a very quick motion that causes light, as well as a more robust motion that causes heat.
    b. The motion must be vibrational in nature. He posits this after ruling out circular, *turbinated* or any other irregular motion.
    c. The vibration is very short.
2. As for how light propagates, Hooke provides the concept of a certain medium and underlines its properties [24, pp. 56-57].
    a. It is a body susceptible and impartible of this motion that will deserve the name of a transparent.
    b. It must be homogeneous.
    c. The motion and constitution of the parts must be such, that the pulse of the luminous body may be propagated through it to the greatest imaginable distance in the least imaginable time. He then says that it may not be in an instant, as he had no experimental methods to confirm.
    d. The rays of light are propagated through this homogeneous medium in straight lines like that of rays emanating from the centre of a sphere.
3. He then suggests vibrational 'wavefronts' that occur perpendicular to the direction of propagation (Schem. VI, Fig. 1) [24].
4. The change of angle in the same wavefronts post-refraction is what causes the regions of red and blue around the shadow as observed in Grimaldi's observations.

His ideas were developed on the basis of trying to find a theory of colours and were highly philosophical and had a lack of mathematical analysis.

Along with Hooke, Ignace-Gaston Pardies, also suggested a wave theory of light but unlike Hooke, who assumes the Cartesian sine ratio (Eq. 12), Pardies uses the Fermatian ratio (Eq. 14) to explain refraction. Pardies had also perfomed demonstrations on Fermat's principle of least time [8].

Wave theory itself, is perhaps best described by the works of Christian Huygens. Huygens' work, the *Traité de la Lumière*, was understood as an attempt to fix Descartes' theory and solve its inaccuracies.

I shall try to put forward a thorough summary of the nature of light discussed in this 138-page book.

In Chapter I of [25], Huygens underlines that light consists in the motion of some sort of matter. This motion is rapid, which he explains by suggesting examples of fires and how light collected at a focus starts fire.

He compares the propagation of light to sound by suggesting that if light takes time for its passage, it propagates via successive motions – like that of several balls together generating a pulse when a force is applied on one of them – on the intervening matter like sound does, by spherical surfaces and waves. He calls them waves as he likens them to ripples formed after throwing a stone in water.

Then, he tries to determine if light has a finite velocity. He suggests a case study on why it may seem that the propagation of light is instantaneous using Descartes' lunar eclipse idea and then cites Rømer's work to suggest that light is finite and is about 100,000-600,000 times faster in contrast to sound!

---

[19] Da Vinci's works are also said to have contained hints on diffraction but Grimaldi is credited for explaining the phenomenon as it is.

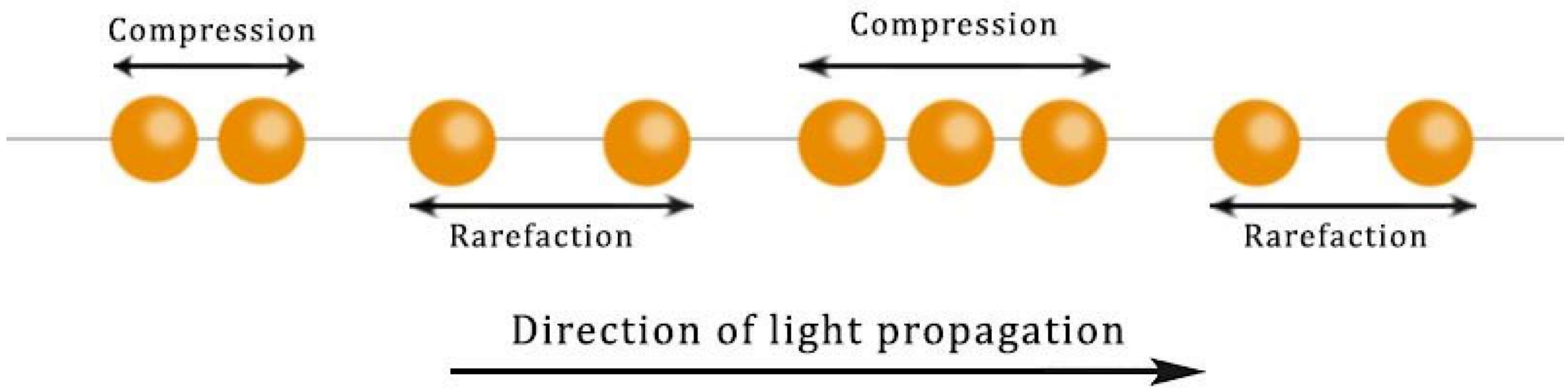

*Figure 3: Huygens' idea of light waves interacting with aether particles. He posits the process is similar to that of sound but can occur in absence of air as well.*

He outlines the differences between the 'propagators' of light and sound by suggesting that the medium of light is different to that of sound, and that light can travel even without the support of air. He calls this medium '*aethereal matter*'. Huygens posits that the movement in the aethereal medium is elastic and the 'springiness' such that, the velocity of light stays the same (Figure 3).

An example of a candle is provided in [25, pp. 16-17] where he does entertain the idea of interference, allowing for 'crashing of light waves' and suggests that aether medium has the capacity to transmit multiple waves at once. He uses both ideas to show how waves travel from celestial distances like that of the sun,

"But what may at first appear full strange and even incredible is that the undulations produced by such small movements and corpuscles, should spread to such immense distances; as for example from the Sun or from the Stars to us. For the force of these waves must grow feeble in proportion as they move away from their origin, so that the action of each one in particular will without doubt become incapable of making itself felt to our sight. But one will cease to be astonished by considering how at a great distance from the luminous body an infinitude of waves, though they have issued from different points of this body, unite together in such a way that they sensibly compose one single wave only, which, consequently, ought to have enough force to make itself felt. Thus this infinite number of waves which originate at the same instant from all points of a fixed star, big it may be as the Sun, make practically only one single wave which may well have force enough to produce an impression on our eyes." [25, p. 18]

Huygens furthers postulates each particle on the aether of an object generates secondary wavelets of its own and forms a new wavefront with the progression of time. This is known as the *Huygens' Principle*.

For reflection, he uses the same geometrical method by applying it to wavefronts. He preserves for Laws of Reflection for all types of surfaces and shows that it is not necessary for a flat surface for reflection.

For the phenomenon surrounding refraction, he hypothesises the following:

1. Aethereal matter passes through all transparent bodies.
2. Particles of transparent bodies have a very small recoil and the progression of the waves of light will be slower in the interior of such bodies than it is outside in the aethereal matter.
3. Opaque objects do not transmit incoming light waves onto surrounding outer aether particles.
4. Metals are given as examples of substances that are really opaque, and are said to contain 'soft particles' along with a few 'hard particles' (those that cause reflection and refraction) to allow for the phenomenon of diffuse reflection and refraction.

Huygens mentions Fermat's principle as a fundamental property of waves and uses it to define how one can arrive at Snell's Law[20] [25] [26].

[20] I have, here, provided a method using mathematical representations. The original text [25, pp. 43-45] uses a more textual approach.

We consider a wave entering a denser media from a rarer media,

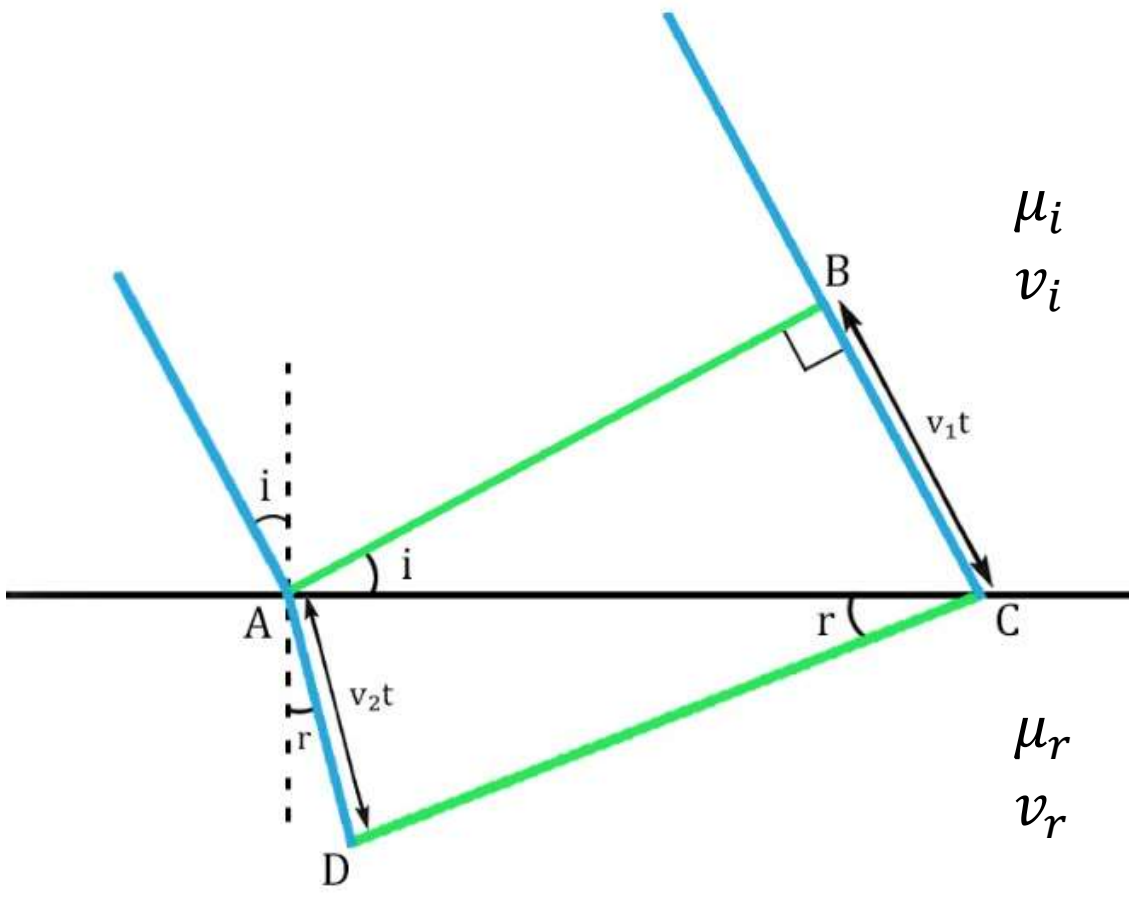

*Figure 4: A light wave with wavefront **AB** and wavelength **BC** in a rarer media refracts with wavefront **DC** and wavelength **AD** in the denser media.*

From Figure 4,

$$BC = v_i t = \lambda_i \quad (45a)$$

$$AD = v_r t = \lambda_r \quad (45b)$$

Now,

$$\sin i = \frac{BC}{AC} = \frac{v_i t}{AC} \quad (46a)$$

$$\sin r = \frac{AD}{AC} = \frac{v_r t}{AC} \quad (46b)$$

Dividing Eqs. 46a and 46b, we return, once again to Eq. 14. This shows proof that light propagates with slower speeds of light in denser media in agreement with the Fermatian ratio.

Similarly, BC and AD can be represented via their corresponding wavelengths.

$$\frac{\lambda_i}{\lambda_r} = \frac{v_i}{v_r} \quad (47)$$

Now, Eq. 47 shows a constant ratio of wavelengths (via Eq. 14). And if we rearrange,

$$\frac{v_i}{\lambda_i} = \frac{v_r}{\lambda_r} = f \quad (48)$$

This suggests that the frequency of a light wave is constant in refraction[21]. Both the wavelength $\lambda$ and velocity $v$ reduce when entering denser media [26].

The rest of the book deals with special cases of atmospheric refraction and double refraction in Iceland Crystal. He also provides a discussion on refraction and reflection in transparent bodies.

After Huygens, Leonhard Euler put forward his own theory of light.

Euler's theory has its roots on the velocity and refrangibility of light and its colours. He had two guiding principles [19]:

1. Light is a wave transmitted through an elastic aether like sound is propagated in air.
2. Frequencies are a fundamental concept; each colour has its own proper frequency.

Euler propounds that whenever light of a particular colour with a velocity $v_0$ propagates via the aether, it generates pulses with the same distance, let us suppose, $\lambda$. This gives us an equation,

$$\frac{v_0}{\nu} = \lambda \quad (49)$$

$\nu$ represents the frequency of that colour in aether medium. It is understandable that $v_0$ is the same for all colours.

Euler, in line with Huygens, suggests that these pulses move slower in a denser medium than in a rarer one. He also suggests that successive pulses are dependent on each other.

$$v = v_0 + \beta \frac{\nu}{v_0} \quad (50)$$

Which we can simplify (via Eq. 49) to,

$$v - v_0 = \frac{\beta}{\lambda} \quad (51)$$

$v$ is the velocity of a sequence of pulses and $\beta$ is a positive constant so that the velocity of a sequence of pulses is increased compared to the velocity of a single pulse [19].

If we apply this idea to Snell's Law considering a medium interface $i = 1{,}2$.

$$\mu_{21} = \frac{v_1}{v_2} = \frac{v_{0,1} + \beta \frac{\nu}{v_{0,1}}}{v_{0,2} + \beta \frac{\nu}{v_{0,2}}} \quad (52)$$

$v_{0,1}$ and $v_{0,2}$ are the corresponding pulse velocity for the aether in media 1 and 2.

Empirically, the term $\beta\, {}^{\nu}/_{v_0}$ is small, so by adjustment and approximation of terms, we get,

$$\mu_{21} = \frac{v_{0,1}^3 \cdot v_{0,2}}{\left(v_{0,2} \cdot v_{0,1}\right)^2 + \beta\nu\left(v_{0,1}^2 - v_{0,2}^2\right)} \quad (53)$$

This equation basically underlines a relation between the frequency and refractive index.

[21] This is also self-evident from the fact that the time taken by the light wave in both media is the same via Fermat's principle and $1/t = f$.

Euler uses this to find out that red has the highest frequency among colours and violet the least.

In the following years, Euler provides a new relation for Eq. 50, suggesting that,

$$v = f(v_0, v) = v_0^{1+F(v)} \qquad (54)$$

$F(v)$ acts a linear function of frequency. If we consider a case a single pulse exists and $v = 0$ then the function gives $F(0) = 0$. This suggests that $v = v_0$. Another case where we consider $v_0 = 1$ gives us the velocity of light $c$ as,

$$c = f(1, v) = 1 \qquad (55)$$

This suggests that the speed of light is independent of its frequency. Euler posits that the aether is the only medium capable of allowing this[22]. Using Eq. 54, we can test for Newton's test of any colour changes in any of Jupiter's moons. Euler, in fact did test for this, and confirmed that there was no observable colour change.

Pedersen concludes his analysis by saying that, "Although the colour change of the satellites was widely discussed at the time by scientists, nobody seems to have commented on Euler's explanation. Euler's theory was, however, related to observations and experiments." [19]

Thomas Young, the man who along with Helmholtz developed the trichromatic theory of vision, was also a believer in the wave theory in the landscape of Newtonian classical physics of England.

In the 1801 Bakerian Lecture, Young opens by suggesting inconsistencies in Newton's theory. He puts forward the following propositions [27],

1. All impulses are propagated in a homogeneous elastic medium with an equable velocity.
2. An Undulation (forward-moving pulse) conceived to originate from the Vibration of a single Particle, must expand through a homogeneous Medium in a spherical Form, but with different quantities of Motion in different Parts.
3. A Portion of a spherical Undulation admitted through an Aperture into a quiescent Medium, will proceed to he further propagated rectilinearly in concentric Superficies, terminated laterally weak and irregular Portions of newly diverging undulations.
4. When an Undulation arrives at a Surface which is the Limit of Mediums of different Densities, a partial Reflection takes place, proportionate in Force to the Difference of the Densities.
5. When an Undulation is transmitted through a Surface terminating different Mediums, it proceeds in such a of the Angles of Incidence and Refraction are the constant Ratio of the Velocity of Propagation.
6. When an Undulation falls on the Surface of a rarer Medium, so obliquely that it cannot be regularly refracted, it is totally reflected, at an Angle equal to that of its Incidence.
7. If equidistant Undulations be supposed to pass through a Medium, of which the Parts are susceptible of permanent Vibrations somewhat slower than the Undulations, their Velocity will be somewhat lessened by this vibratory Tendency; and, in the same Medium, the more, as the Undulations are more frequent.
8. When two Undulations, from different Origins, coincide either perfectly or very nearly in Direction, their joint effect is a Combination of the Motions belonging to each.

Based on all the eight propositions, Young provides a final proposition:

> *Radiant Light consists in Undulations of the luminiferous aether.*

Young then outlines corollaries on various phenomena relating to reflection, refraction and diffraction.

In corollary 2 [27, pp. 37-41], when discussing the case of colours in thin lenses, Young talks about the spectrum in terms of musical harmonies. He quotes Hooke's theory of light based on thin lenses in [24].

Young also provides a table (Table 1) based on Newton's measures of the thicknesses reflecting the different colours.

We can estimate what Young took to be the speed of light via the equation,

$$c = \nu\lambda \qquad (59)$$

Plugging in data from Table 1 gives us $c$ in the range of 311,700-312,800[23] km/s. However, it is remarkable to note that Young was not off by very much from the modern-day values.

[22] 'This' refers to the property of aether to allow all kinds of different rays to pass through it with the same velocity [19].

[23] This is seen to be off by 4.59 – 4.96% of James Bradley's calculation based on stellar aberrations. See Section 6.

| COLOURS | WAVELENGTH (IN NANOMETERS)[24] | FREQUENCY (IN TERAHERTZ) |
|---|---|---|
| **EXTREME (R)** | 675.6 nm | 463 THz |
| **RED** | 650.2 nm | 482 THz |
| **R-O INTERMEDIATE** | 624.8 nm | 501 THz |
| **ORANGE** | 609.6 nm | 512 THz |
| **O-Y INTERMEDIATE** | 596.9 nm | 523 THz |
| **YELLOW** | 576.6 nm | 542 THz |
| **Y-G INTERMEDIATE** | 556.3 nm | 561 THz |
| **GREEN** | 535.9 nm | 584 THz |
| **G-B INTERMEDIATE** | 515.6 nm | 607 THz |
| **BLUE** | 497.8 nm | 629 THz |
| **B-I INTERMEDIATE** | 480.1 nm | 652 THz |
| **INDIGO** | 469.9 nm | 665 THz |
| **I-V INTERMEDIATE** | 459.7 nm | 680 THz |
| **VIOLET** | 442.0 nm | 707 THz |
| **EXTREME (V)** | 424.2 nm | 735 THz |

*Table 1: Young's table of colours. He outlines the length of their undulations (wavelengths) and their frequencies [27, p. 39].*

Young also talks about the phenomena of wave interference (in-line with Huygens) like the phenomena of constructive and destructive interferences.

We consider two waves $L_1$ and $L_2$, then by the superposition principle we get [26],

$$L_1 = A\cos\omega t \qquad (57a)$$

$$L_2 = A\cos(\omega t + \phi) \qquad (57b)$$

Therefore,

$$L = L_1 + L_2 = 2A\cos\omega t \cos\frac{\phi}{2} \qquad (58)$$

This suggests that whenever the phase difference $\phi$ between $L_1$ and $L_2$ will be in even multiples of $\pi$, light will be of maximum intensity (constructive) and minimum (destructive) for odd multiples.

Young devised an experiment in the 1803 Bakerian Lecture, which confirmed interference among light waves mathematically via a simple experiment [28].

The experiment is defined by Young as such,

"I made a small hole in a window-shutter, and covered it with a piece of thick paper, which I perforated with a fine needle. For greater convenience of observation, I placed a small looking glass without the window-shutter, in such a position as to reflect the sun's light, in a direction nearly horizontal, upon the opposite wall, and to cause the cone of diverging light to pass over a table, on which were several little screens of card-paper. I brought into the sunbeam a slip of card, about one-thirtieth of an inch in breadth, and observed its shadow, either on the wall, or on other cards held at different distances." [28, p. 2]

The slip of card 'split' the oncoming sunlight into two rays, and generated an alternating pattern of dark and light patterns. This became the basis of what came to be known as the *Double Slit Experiment.*[25]

Via Fig. 10.12b in [26], the positional differences in the maxima and minima of the light intensity from the split rays come as,

$$S_2P - S_1P = n\lambda \qquad (59)$$

Maxima will appear provided that $n \in \mathbb{W}$. Now, via geometry,

$$(S_1P)^2 = D^2 + \left(x - \frac{d}{2}\right)^2 \qquad (60a)$$

$$(S_2P)^2 = D^2 + \left(x + \frac{d}{2}\right)^2 \qquad (60b)$$

Now, subtracting Eq. 64 from Eq. 65,

$$(S_2P)^2 - (S_1P)^2 = 2xd \qquad (61)$$

Or,

$$S_2P - S_1P = \frac{2xd}{S_2P + S_1P} \qquad (62)$$

Given, $x, d << D$, we can approximate Eq. 62 as,

$$S_2P - S_1P = \frac{2xd}{2D} = \frac{xd}{D} \qquad (63)$$

Equating Eqs. 59 and 63,

[24] The values were provided in inches in the original document [27, p. 39]. They have been converted to nanometres for ease.

[25] It must be noted that the experiment mentioned in the journal never uses double slits. Although, Young does suggest that double slits would also have the same effect.

$$n\lambda = \frac{xd}{D} \Rightarrow x = \frac{n\lambda D}{d} \quad (64)$$

The distance between two consecutive fringes are given by,

$$x_{n+1} - x_n = \frac{\lambda D}{d} = \beta \quad (65)$$

Here, $\beta$ is the fringe width. Bright spots appear for multiples of $\beta$ and dark spots appear for half-odd multiples of $\beta$. Similarly, the angle is given by,

$$\sin\theta_1 = \frac{n\lambda}{d} \quad (66)$$

The special optical properties of calcite[26] is re-introduced onto the scene by Étienne-Louis Malus, who studies light rays reflected off a non-metallic surface. He notices that at an appropriate angle, it behaves like one of the two rays emerging from a calcite crystal. He calls this *polarisation.* The plane on which the reflection takes place is known as the *plane of polarisation.*

He tried to explain the phenomenon in corpuscular terms – he was a follower of Laplace – but in modern context, it is seen that polarisation is a fundamental property of waves.

He devised a law to relate intensities between two successive polarisers given by the following formula,

$$I = I_0 \cos^2 i \quad (67)$$

Here, $I_0$ and $I$ are the intensities of the source and second polariser respectively.

We consider a general transmission coefficient[27],

$$\frac{I}{I_0} = \frac{1}{2} \quad (68)$$

A few years later, in 1815, David Brewster finds a relation between the refractive index and the Brewster's angles (polarisation angles) as follows,

He posits in proposition IV of [29],

$$i + r = 90° \quad (69)$$

Now, applying Snell's Law and using the relation in Eq. 69,

$$\sin i = \mu_{21}\sin(90 - i) \quad (70)$$

Or,

$$\sin i = \mu_{21}\cos i$$
$$\Rightarrow \tan i = \mu_{21} \quad (71)$$

Now, if the angle of incidence is the polarising angle $\theta_B$,

$$\tan\theta_B = \mu_{21} \quad (72)$$

By inversion,

$$\theta_B = \arctan\mu_{21} \quad (73)$$

This relation is also known as the Brewster's Law.

Influenced by the likes of Hooke and Huygens, a French engineer, Augustin-Jean Fresnel provided his own theory of light.

Fresnel started to study interference and polarisation by independently recreating many of the experiments by Huygens, Hooke, *et al.*

The first memoir was on the diffraction and dealt with Newton's inaccuracies, Young's experiments and the wave theory. Fresnel used wave equations to define interference and also applied the Huygens' Principle to explain the phenomena of diffraction [30].

By consequence of the principle, each secondary wavelet would generate backward waves, which Huygens doesn't account for. The tangential representation of a wavefront, showed that it was the same irrespective of wavelength. This became problematic as the principle didn't agree with the mathematical equations (Eqs. 70 and 71) and also experimental evidence on diffraction.

Fresnel suggested that the interference and the overlapping regions is what led to the formation of the wavelet [30]. This then became known as the *Huygens-Fresnel Principle.*

Fresnel assumes two waves of light in the form of a general wave equation [31],

$$W_1 = W_2 = A\sin(\omega t - kx) \quad (74)$$

Now, if the second wave traverses an additional distance than the prior, then the modified equation becomes,

$$W_2 = A\sin\big(\omega t - k(x + d)\big) \quad (75)$$

Where, the term $kd$ corresponds to a phase $i$,

$$i = kd = \frac{2\pi d}{\lambda} \quad (76)$$

It is seen by Fresnel (mathematically) that the wave in Eq. 75 can be spilt into two component waves with amplitudes $A\sin i$ and $A\cos i$, suggesting that their phase difference would be 90°.

This suggests that the resultant amplitude of the two waves would be square root of the sum of the squares of the component amplitudes similar to the sum of two vectors.

It is here in its application, that we see Fresnel's diffraction[28] take place.

---

[26] See Chapter V of [25].

[27] The polarization angles come in all possibilities in natural light and hence an average value, $\cos^2 45° = 1/2$ is taken. This basically suggests that the polarised intensity is always between 0 – 50% of the incident intensity of al light wave. Polarisation doesn't occur beyond this limit.

[28] This is the near-field approximation of the Kirchhoff-Fresnel diffraction. Fraunhofer diffractions apply to far-fields.

Fresnel considers an opaque body, which provides part hindrance to light propagation.

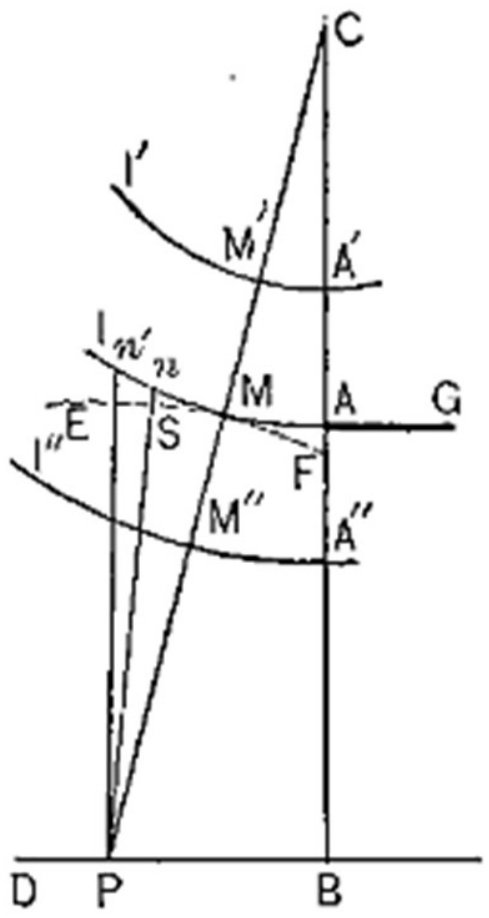

*Figure 5: Fresnel's idea of how light waves emanating from **C** behave with an opaque object **AG**. [30, p. 119]*

From Figure 5, we can compute the distance of $nS$ as,

$$nS = \frac{z^2(a+b)}{2ab} \qquad (77)$$

This is the distance between the wavelets AMI and EMF. Here, $z$ is equal to $Mn$ and $a$, $b$ correspond to $CA$ and $AB$. Fresnel approximates this to hold true for all points close to $M$[29] [30] [31].

If we consider such a wave coming from the source $M$ with a fixed wavelength $\lambda$. The sum of all such wavelets per element $nn' = dz$, for the sine component comes via Eqs. 76 and 77 as,

$$A_1 = \int \sin\left(\pi \frac{z^2(a+b)}{ab\lambda}\right) dz \qquad (78a)$$

And for the other component [30],

$$A_2 = \int \cos\left(\pi \frac{z^2(a+b)}{ab\lambda}\right) dz \qquad (78b)$$

If we are to simplify these, we get the normalised Fresnel integrals[30], provided we give a limit. Fresnel does suggest a limit of $A$ to $M$ in his texts [30].

Using Eqs. 78a and 78b, The continuous sum (integral) for the resultant amplitude of all wavelets comes,

$$A = \sqrt{A_1^2 + A_2^2} \qquad (79)$$

And the intensity comes as,

$$I = A_1^2 + A_2^2 \qquad (80)$$

Fresnel's results on diffraction agreed with the experiments he carried out. It is seen that the intensity falls rather swiftly as we move to areas closer to the geometric shadow [30, pp. 129-131].

Fresnel summarised the new interference-based principle as [30, p. 144],

"…light is appreciable only in those points where the secondary waves coincide (in phase) exactly. […] absence of luminous disturbance can occur only when the secondary waves are in (direct) opposition."

When the committee (consisting of Biot, Laplace, Gay-Lussac, Arago and Poisson), was deliberating over the memoir, Siméon Denis Poisson postulated that by using Fresnel's principle, a circular object illuminated by a point-source, would show a bright spot in the centre of the shadow, illuminated as brightly as the outer bounds of the shadow. This became known as Poisson's/Arago's spot. This absurd consequence would be the make-or-break for Fresnel.

Arago performed such an experiment and succeeded in generating a bright spot at the centre of a shadow.

Fresnel then turns to polarisation, an important phenomenon in optics during the time. His works with a birefringent crystal shows an anomalous double-slit interference pattern. Further work by Arago in a more general experiment, gives rise to the *Fresnel-Arago laws*.

1. Two orthogonal, coherent linearly polarised waves cannot interfere.
2. Two parallel coherent linearly polarised waves will interfere like natural light.
3. The two-constituent orthogonal linearly polarised states of natural light cannot

---

[29] If we consider the wavelet generated from M to have a path length $x$, and the others are waves differing by some distance $d$ (here equal to $\lambda/4$ [30]), then we can apply the theory in the sections surrounding Eqs. 75 and 76.

[30] Normalised Fresnel integrals come in the form,

$$S(x) = \int_0^x \sin(qv^2)dx$$

$$C(x) = \int_0^x \cos(qv^2)dx$$

Fresnel did use q (from 'quadrant') to represent $\pi/2$ and $v^2$ is a simplification of the terms in Eqs. 78 and 79 [30, p. 121].

interfere to form a readily observable interference pattern, even if rotated into alignment (because they are incoherent).

Fresnel also noticed that if polarised light passed through a crystal lamina, it was split into ordinary and extraordinary waves, and these were perpendicularly polarised and therefore did not interfere due to Law 3. Post-alignment, however, it was seen that images with complementary hues were formed following the relation,

$$I_O + I_E = 1 \tag{81}$$

Where, $I_O$ and $I_E$ correspond to the ordinary and extraordinary ray intensities.

André-Marie Ampère had suggested to Fresnel that polarisation could be explained if the polarised waves were partly transverse [32].

If we consider Law 3 of the Fresnel-Arago Laws, we see that it is only possible if light was completely transverse.

Young also hinted at polarised light being transverse and considered that Malus' Law was only possible if light was a transverse wave.

If we consider light to be a transverse wave, we require the aether to be a solid. But it is seen that the aether does not interact with other bodies. So, Augustus-Louis Cauchy suggests that there must be some sort of a drag and that the ether had negative compressibility.

This made aberrations hard to define and negative compressibility made the fluid unstable. Stokes then suggest that the aether could be fluid like in slower speeds and solid like in faster speeds.

Cauchy also determines a relation between the refractive index and wavelength of a colour,

$$\mu(\lambda) = A + \frac{B}{\lambda^2} + \frac{C}{\lambda^4} + \cdots \tag{82}$$

We use only the first two terms of this equation for general purposes, were $A$, $B$ and $C$ are empirical constants.

## 5. LIGHT AND ELECTROMAGNETISM

The history of light and EM Waves being similar start with the hypothesis that light is a transverse wave.

The first notable linkage between both light and EM waves was by Michael Faraday. Faraday had used various materials (Entries 7484-7496) in an electrical apparatus through which he passed a beam of polarised light. He observed there were no notable effects.

He then tried to apply a similar idea with magnetic forces, where he passed 'lines of magnetic force'[31] and a beam of polarised light through various materials. He did not have conclusive results with air, flint glass, rock crystal and calcareous spar (Entries 7499-7503). In heavy glass (7504), however, he observes depolarisation of the polarised beam. He suggests that magnetic forces and light have some relation between them [33]. Further detailed experiments lead to the *Faraday Effect* and the *Faraday Rotation*. He later proposed that light itself is a high-frequency electromagnetic vibration, which could propagate even in the absence of a medium.

In 1856, Wilhelm Eduard Weber and Rudolf Kohlrausch working on electrodynamics gave the following expression for the fundamental law of electrical action[32] in their paper [34] [35],

$$\frac{qq'}{r^2} - \frac{qq'}{r^2}\frac{1}{C_w^2}\left[\left(\frac{dr}{dt}\right)^2 - 2r\frac{d^2r}{dt^2}\right] \tag{83}$$

Here, we see the constant $C_w$, a unit of velocity. The determination of which was done by the following,

The ratio of the magnetic unit of current intensity to the mechanical unit came as,

$$\frac{E}{2\tau} : 1 = 155{,}370 : 1 \tag{84a}$$

For magnetic unit of current intensity to the electrodynamic unit,

$$\sqrt{2} : 1 \tag{84b}$$

For electrodynamic unit for the current intensity to the mechanical one,

$$C_w : 4 \tag{84c}$$

Combining all,

$$C_w = \sqrt{2}\frac{E}{\tau} = \sqrt{2} \times 310{,}740\, km/s \tag{85}$$

[31] This is Faraday's idea of magnetic fields. We will see this terminology being used in Maxwell's papers as well.

[32] The law basically states that the force exerted is not only dependent on the Coulombic parameters in Gaussian units, but also on the relative velocity and acceleration. From Eq. 84, we can break it into two parts:

$$F_c = \frac{qq'}{r^2}$$

$$F_{ed} = -\frac{qq'}{r^2}\frac{1}{C_w^2}\left[\left(\frac{dr}{dt}\right)^2 - 2r\frac{d^2r}{dt^2}\right]$$

Here, the first equation (Coulombic) is for the electrostatic force and the second one is the electrodynamic force.

Here, $C_w$ became the Weber's constant. It was interesting to note that the value of $E/\tau$ was so close to the speed of light (as measured by Fizeau in 1849 to be 315,000 km/s and Foucault in 1862 to be 298,000 km/s). This suggested a possible linkage between the fields of optics and electrodynamics, but Kirchhoff and Weber apparently missed this.

The following year, Gustav Kirchhoff in [36] shows that if the resistance of a wire tends to zero (ideal wire), the electricity in the wire progresses like a wave in a taught string with the velocity of light in empty space based on previous experiments taking note of the coincidence.

In the start of the next decade, James Clerk Maxwell revolutionises electromagnetism by his famous Maxwell's Equations[33],

$$\nabla\cdot\mathbf{E}=\frac{\rho}{\varepsilon_0} \qquad (86a)$$

$$\nabla\cdot\mathbf{B}=0 \qquad (86b)$$

$$\nabla\times\mathbf{E}=-\frac{\partial\mathbf{B}}{\partial t} \qquad (86c)$$

$$\nabla\times\mathbf{B}=\mu_0\left(\mathbf{J}+\varepsilon_0\frac{\partial\mathbf{B}}{\partial t}\right) \qquad (86d)$$

If we consider a free space with no current ($\mathbf{J}=0$) or charge ($\rho=0$), Eqs. 86a and 86d reduce to,

$$\nabla\cdot\mathbf{E}=0 \qquad (87a)$$

$$\nabla\times\mathbf{B}=\mu_0\varepsilon_0\frac{\partial\mathbf{B}}{\partial t} \qquad (87b)$$

Now, if we take the curl of Eqs. 86c and 87b and $1/\sqrt{\mu_0\varepsilon_0}=c$, we get,

$$\frac{1}{c^2}\frac{\partial^2\mathbf{E}}{\partial t^2}-\nabla^2\mathbf{E}=0 \qquad (88a)$$

$$\frac{1}{c^2}\frac{\partial^2\mathbf{B}}{\partial t^2}-\nabla^2\mathbf{B}=0 \qquad (88b)$$

These come in the form of standard wave equations[34], showing that the phase velocity of EM waves is equal to the speed of light.

In Part III of his paper, *On Physical Lines of Force* [37, p. 22], Faraday himself tries to find the speed at which EM waves travel. He shows the relation[35],

$$\pi m=V^2\mu \qquad (89)$$

Here, $m$ is referred to as the 'coefficient of transverse elasticity' and $\mu$ is the 'coefficient of magnetic induction'. $E$ is described as "…the number by which the electrodynamic measure of any quantity of electricity must be multiplied to obtain its electrostatic measure." [37, p. 21]

This is simplified further by the relation,

$$E^2=\pi m \qquad (90)$$

Which holds true for media whose elasticity depends entirely upon forces acting between pairs of particles, and the ratio of the forces is in 6:5 [37, p. 18]. The media that Maxwell considers here, is part aether and part matter.

Now, combining Eqs. 90 and 91,

$$E=V\sqrt{\mu} \qquad (91)$$

For air or vacuum, $\mu$ is 1 and therefore $E=V$. Using Weber and Kohlrausch, he calculates V as 310,740 km/s and writes the following,

"The velocity of transverse undulations in our hypothetical medium, calculated from the electro-magnetic experiments of MM. Kohlrausch and Weber, agrees so exactly with the velocity of light calculated from the optical experiments of M. Fizeau, that we can scarcely avoid the inference that *light consists in the transverse undulations of the same medium which is the cause of electric and magnetic phenomena*." [37, p. 22]

A similar method is also used in [38], and Maxwell shows that his experiment never used light as a part and Foucault's did not use electricity and magnetism. Hence, the coincidence was too much in agreement of the fact that light and EM waves are the same thing.

Heinrich Rudolf Hertz later experimented on Maxwell's equations and succeeded in generating radio waves that behaved like light – showing optical phenomena like reflection, refraction, *etc.* This confirmed the existence of EM waves.

---

[33] Given in the differential SI (Modern Maxwell-Heaviside) form.

[34] For a wave with its electric field along the x-axis, magnetic field along the y-axis and direction of propagation along the z axis, we can also show the wave equations as,

$$E_x=E_0\sin(kz-\omega t)$$

$$B_y=B_0\sin(kz-\omega t)$$

It must be noted that $c=\omega/k$ as $\omega=2\pi\nu$ and $k=2\pi/\lambda$. The relation between the amplitudes of the electric and magnetic fields come as, $E_0/B_0=c$ [26] [39]. For the general case, we have the vector forms,

$$\mathbf{E}(\mathbf{r},t)=\mathbf{E_0}\sin(\mathbf{k}\cdot\mathbf{r}-\omega t)$$

$$\mathbf{B}(\mathbf{r},t)=\mathbf{B_0}\sin(\mathbf{k}\cdot\mathbf{r}-\omega t)$$

$\mathbf{k}$ is the wave vector analogous to the scalar $k$. $\mathbf{r}$ is the position vector and $\mathbf{E_0}$, $\mathbf{B_0}$ are the amplitude vectors of the EM fields.

[35] From Newton's isothermal equation for the speed of sound.

As a result of the EM revolution, polarisation gained an entirely new definition, rehauling almost all the known physics for it at that time. Polarisation became the phenomena involving the realignment of the EM fields in unpolarised (randomised planes of vibrations) waves to have a fixed field plane, thereby having a fixed direction for the other field and direction of propagation [26].

The Fresnel-Arago Laws can also be explained via electric fields. We consider two wave equations for the electrical components,

$$\mathbf{E_1}(\mathbf{r}, t) = \mathbf{E_{01}} \cos(\mathbf{k_1} \cdot \mathbf{r} - \omega t + \epsilon_1) \quad (92a)$$

$$\mathbf{E_2}(\mathbf{r}, t) = \mathbf{E_{02}} \cos(\mathbf{k_2} \cdot \mathbf{r} - \omega t + \epsilon_2) \quad (92b)$$

The intensity is given as,

$$I = \frac{1}{2}\varepsilon_0 \mathbf{E}^2 c = \frac{1}{2} c \cdot \mathbf{E_{01}} \cdot \mathbf{E_{02}} \cos\delta \quad (93)$$

The phase difference $\delta$ is given as,

$$\delta = \mathbf{r}(\mathbf{k_1} - \mathbf{k_2}) + \epsilon_1 - \epsilon_2 \quad (94)$$

Using Eq. 93, we see that the laws hold true.

Very interestingly, Feynman shows in [39] that refractive index originates from interactions in the electric field. He shows that the refractive index is dependent on the properties of atoms of the material and the light frequency,

$$\mu = 1 + \frac{Ne^2}{2\varepsilon_0 m_e(\omega_0^2 - \omega^2)} \quad (95)$$

$N$ represents the number of charges per unit volume in a refracting plate, $\omega_0$ is the resonant frequency of the electron in an atom and $\omega$ is the angular frequency of light.

He then improves Eq. 95 based on an atom having several resonant frequencies,

$$\mu = 1 + \frac{e^2}{2\varepsilon_0 m_e} \sum_k \frac{N_k}{(\omega_0^2 - \omega^2 - i\gamma_k\omega)} \quad (96)$$

Here, $i\gamma_k\omega$ is introduced as a damping factor[36]. This actually suggests that the refractive index of a material is complex in nature,

$$\mu = \mu' + ik \quad (97)$$

In most cases, we just use the real part $\mu'$ and $k$ is given as the extinction ratio.

## 6. ABERRATIONS AND AETHER DRAG

The phenomenon of 'aberration[37]' started in search of stellar parallaxes. Several people (like Hooke, Flamsteed, *et al.*) noted deviations in stellar positions across the year.

In 1725, James Bradley and Samuel Molyneux attempted to answer this. They collected data from a telescope which showed that the star γ Draconis passed more southerly (by 39 arsecs) than it should, contrary to the motion of the annual parallax [40].

Bradley had first considered nutation[38], but saw that it did not hold true for all stars. He observed 35-Camelopardalis, a 'small star' almost opposite in right ascension to γ Draconis, at about the same distance from the North Pole of the Equator and noticed that the declination was half of γ Draconis passing northerly (by 19 arsecs).

He considers experimental error and refraction but doesn't find them to be suitable. He then states the reason as,

"…if Light was propagated in Time, the apparent Place of a fixt Object would not be the same when the Eye is at Rest, as when it is moving in any other Direction, than that of the Line passing through the Eye and Object; and thar, when the Eye is moving in different Directions, the apparent Place of the Object would be different." [40, p. 646]

Bradley finds from the data of several stars to find that the mean of the maximum deviation comes as 40.4 arsecs. He suggests that this explains[39] the pole of the elliptic that formed as a result of the elliptical orbit of the Earth.

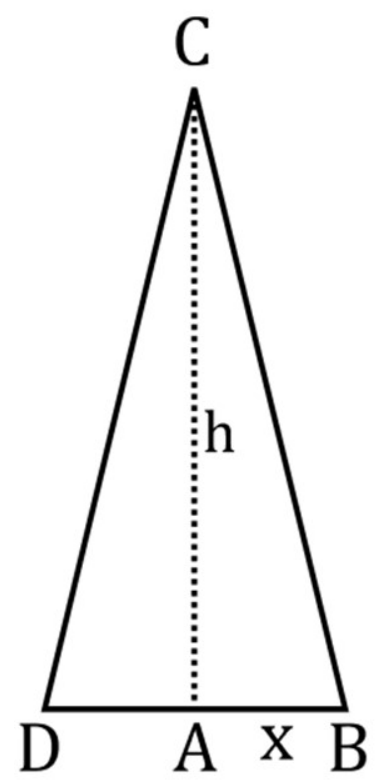

*Figure 6: Bradley's description of a star at point* ***C*** *and the motion of the observer along* ***DB****.*

From Figure 6, the distance $AC$ comes as,

$$AC = ct = h \quad (98)$$

[36] The system is considered to be a damped oscillator. $\gamma_k$ is the damping coefficient.

[37] Here, aberrations refer to 'stellar aberrations' in celestial objects. Chromatic aberrations are discussed briefly in the earlier sections.

[38] Change in the orientation of the axis of rotation of a spinning astronomical object. For earth, the causes of nutation are the gravitational fields of other celestial bodies (especially the Moon).

[39] It is interesting to note that γ Draconis, the star central to the observations, which gave the value of the aberration constant and was ideal for the effect, was chosen Bradley, *et al.* among so many other stars.

This gives us $t = h/c$, making $AB$ as,

$$AB = vt = \frac{vh}{c} = x \tag{99}$$

Here, $v$ is the speed of the eye, and hence, the earth. Therefore, the tangent comes as,

$$\tan\phi \cong \phi = \frac{AB}{AC} = \frac{v}{c} \tag{100}$$

The *constant of aberration* is taken to be half of the mean, as 20.2 arsecs, which gives us a ratio of,

$$\frac{v}{c} = \frac{1}{10210} \tag{101}$$

Bradley calculates via Eq. 99 the speed of light to be in the neighbourhood of 300,000 km/s.

Young had suggested that the aether must pass through the Earth and be immobile to the Sun's reference frame. Young's ideas led to a revision of the generalised angle form of Eq. 100,

$$\tan\phi = \frac{\sin\theta}{{}^{\mu v}/_{c} + \cos\theta} \tag{102}$$

But no such dependence on the refractive index was found after experiments by Arago and Airy.

Fresnel looked at this problem with a wave-view for the first time. He gives a new equation to explain the *aether drag* in glass prisms [41],

$$c_g = \frac{c}{\mu} + v\left(1 - \frac{1}{\mu^2}\right) \tag{103}$$

This held true in experiments conducted by Fizeau.

Stokes's consideration of the aether to be a dilatant made the aether to be completely dragged within and in the vicinity of matter, partially dragged at larger distances, and stays at rest in free space. He provides a mathematical formulation of his aether drag theory in [41] [42],

For an aether stream with velocity dependent on $u$, $v$ and $w$, the equation of the wave front is given as,

$$z = ct + \zeta + C \tag{104}$$

Here, $\zeta = f(x, y, t)$ and is a small quantity. $z$ is the propagation of light along the z-axis.

From further simplification,

$$z = ct + \zeta + (w + c)dt + C \tag{105$a$}$$

This is found to be also equal to,

$$z = ct + \zeta + \left(\frac{d\zeta}{dt} + c\right)dt + C \tag{105$b$}$$

Comparing both Eqs. 105a and 105b and the smallness of $\zeta$ gives us,

$$\frac{d\zeta}{dt} = w \Rightarrow \zeta = \int w\,dt = \int w\frac{dz}{c} \tag{106}$$

Therefore, the directional cosines are given as [41, p. 31],

$$\alpha - \frac{\pi}{2} = \frac{1}{c}\int \frac{dw}{dx}dz \tag{107$a$}$$

$$\beta - \frac{\pi}{2} = \frac{1}{c}\int \frac{dw}{dy}dz \tag{107$b$}$$

Stokes then goes on to show that the aether is irrotational and hence, $dw/dx = du/dz$ and $dw/dy = dv/dz$.

$$\alpha_2 - \alpha_1 = \frac{1}{c}\int_{u_1}^{u_2} du = \frac{u_2 - u_1}{c} \tag{108$a$}$$

$$\beta_2 - \beta_1 = \frac{1}{c}\int_{v_1}^{v_2} dv = \frac{v_2 - v_1}{c} \tag{108$b$}$$

Given $\beta_2 - \beta_1 = u_1 = 0$, we have a singular equation for stars with the angle deviation,

$$\alpha_2 - \alpha_1 = \frac{u_2}{c} \tag{109}$$

Which is Bradley's constant of aberration.

Albert Michelson and Edward Morley performed an experiment in the famous *Michelson-Morley Experiment* to test for the Fresnel aether model, existence of the aether and aether wind and the motion of earth in the aether frame[40].

Due to the aether drag, the speed of light would be different in two orthogonal directions. An interferometer would be used to calculate the deviation of the speeds [43] [44].

Michelson and Morley used a very clever way of detecting the deviation, by using fringe shifts. The fringe shift is given as,

$$n = \frac{2d}{\lambda}\left(\frac{v_{eart}}{c}\right)^2 \tag{110}$$

Where, $d$ is the arm length [43] [44]. By plugging in the experimental values, there should have been a shift of $n$ = 0.4. But, experimental results across the year and at various daytimes, yielded all null results. This was the first blow to the aether theory.

## 7. RELATIVITY AND LIGHT

In order to explain the results, Hendrik Lorentz (and also George FitzGerald) suggests that all bodies undergo 'contraction' along the line of motion by a factor of $\gamma$, the Lorentz Factor, which is given as,

[40] Under classical Galilean transformations, the Maxwell equations were not equivalent and changed with each frame. This caused the speed of light to also change with each frame of reference – which it didn't. Hence, a frame with invariancy of the equations was introduced which was fully stationary no matter the reference frame. This is the aether frame [44].

$$\gamma = \frac{1}{\sqrt{1 - \frac{v^2}{c^2}}} \quad (111)$$

Secondly, he introduced the concept of *local time* by extension of the length contraction hypothesis,

$$t' = t - \frac{vx}{c^2} \quad (112)$$

Where, $t$ is the time co-ordinate in the aether rest frame and $t'$ is the co-ordinate for the observer moving in the same frame. Using this, aberration and the Fizeau experiments could be explained.

He then develops a new set of transformations for which, Maxwell's EM Equations remain invariant [45] [46],

$$x' = \gamma(x - vt) \quad (113a)$$

$$y' = y \quad (113b)$$

$$z' = z \quad (113c)$$

$$t' = \gamma\left(t - \frac{vx}{c^2}\right) \quad (113d)$$

Poincaré suggested in his papers that a good electro-optical theory should explain the Fresnel drag coefficient, conserve electricity and magnetism and stay true to Newton's 3rd Law of Motion [47] [48].

He showed that local time was a necessary result of the 'error' that emerged out of synchronization procedures[41] based on light signals. He also discusses the link of this concept to simultaneity.

This also explained the Doppler effect. If we consider the frequencies generated by the source ($\nu_s$) and the observer ($\nu_o$),

$$\nu_o = \sqrt{\frac{c - v}{c + v}}\nu_s \quad (114)$$

Where, a positive velocity results in a 'redshift' of frequencies towards the red-end of a spectrum and a negative velocity generates a 'blueshift'.

Einstein suggests in his 1905 paper, *On the Electrodynamics of the Moving Bodies*, suggests two postulates [49] [50],

1. Principle of Relativity – The laws by which physical processes occur are the same with respect to any system of inertial coordinates.
2. Constancy of Light – In empty space, light propagates at an absolute speed in any system of inertial coordinates.

Postulate 1 modified Newtonian mechanics and Galilean transformations to make them invariant under Lorentzian conditions. However, we are interested in the aspects regarding Postulate 2.

It is seen that for two frames $K(x, y, z, t)$ and $K'(x', y', z', t')$, Newtonian mechanics states that if $K'$ moves relative to $K$ with a velocity $v$,

$$u_x = u'_x + v \quad (115)$$

We can write them as,

$$u_x = \frac{dx}{dt} \quad (116a)$$

$$u'_x = \frac{dx'}{dt'} \quad (116b)$$

Applying Lorentz transformation (Eqs. 113a and 113d),

$$u_x = \frac{\frac{dx'}{dt'} + v}{1 + \frac{v}{c^2}\frac{dx'}{dt'}} \quad (117)$$

Which can be simplified to,

$$u_x = \frac{u'_x + v}{1 + vu'_x/c^2} \quad (118)$$

Let us consider general velocities $u$ and $u'$. Now, if we suppose that $v = c$,

$$u = c\frac{u' + c}{u' + c} = c \quad (119a)$$

Similarly, if $u'_x = c$,

$$u = c\frac{c + v}{c + v} = c \quad (119b)$$

The consequence of both 119a and 119b is that no matter what the two velocities are, it is not possible to exceed $c$.

Another massive consequence of Einsteinian Special Relativity (STR) is that there is no longer a need for aether[42].

In [47], Poincaré had shown that the transformations could be shown as ordinary rotations of the four-dimensional Euclidean sphere,

$$x^2 + y^2 + z^2 + (ict)^2 = k \quad (120)$$

Hermann Minkowski showed that instead of a sphere, the four-vectors formed a vector space which he called the 'world'. Therefore, a 'world-point' was seen to be the common points of two *light-cones: past and future.* A light cone is defined as the path light takes from an event (world-point) across spacetime [51].

---

[41] For any two observers, the observers try to synchronize their clocks by optical signals while moving in the aether. The observers think that they are in rest and will have the same time, but to an external observer in the aether frame, the clocks are unsynchronised.

[42] Einstein's preference of Maxwell over Newton leads to the fact that Lorentzian transformations can hold in vacuum. Lorentz and Poincaré had the tools to observe so, but chose to side with the aether theory.

This suggested that light was the converting factor between space and time.

In General Relativity (GTR), Einstein suggested three tests to test the validity of GTR [49, pp. 113-122],

1. Motion of the Perihelion of Mercury – Related to Newtonian Gravitational flaws.
2. Deflection of a light ray by a Gravitational Field – This suggests that a ray of light will undergo 'curvature of path' in presence of a strong gravitational field[43] (like the Sun).
3. Displacement of Spectral Lines towards the Red – Light undergoes 'redshifts' and 'blueshifts' in presence of a strong gravitational field.

As per the second test, the theoretical angle of deflection of the sun should come as around 1.7 arsecs.

Arthur Eddington, Frank Dyson, *et al.* confirmed the deviation by conducting observations on stellar trajectories near the Sun during a solar eclipse [52].

The third test was a working problem that is different from relativistic effects of the Doppler effect (Eq. 114), which was confirmed by redshifts in 40 Eridani B and Sirius B.

## 8. SCATTERING

In his 1868 paper, John Tyndall describes how light passed from an electric lamp inside a glass tube with vapours, makes it seem blue in colour. He considers these vapours to contain particles to contain a very small fraction of the wavelength of violet light [53]. This phenomenon came to be known as the *Tyndall Effect*.

He found that the polarisation of a light beam when passed through such a 'incipient cloud' was independent of the Brewster's Law. He also suggests presence of water particles in the sky that may contribute to the blueness.

Lord Rayleigh (John William Strutt) explains the working behind the phenomena of blue sky quantitatively. He suggests that the blue colour of light is generated due to the molecules of air [54]. He also relates the intensity to the fourth power of wave length,

$$I \propto \frac{1}{\lambda^4} \tag{121}$$

This is known as Rayleigh's Law. The heavy dependence on wavelength, leads to a preference of blue over red when the sun remains somewhat perpendicular. We can show the scattered intensity from the molecules as,

$$I = I_0 \frac{8\pi^4 N\alpha^2}{\lambda^4 R^2}(1+\cos^2\theta) \tag{122}$$

Here, $N$ is the number of molecules and $\alpha$ is polarisability[44]. $R$ is the distance from the scatterer.

It is seen that Rayleigh scattering requires the particle to be less than 1/10$^{th}$ of the wavelength of light. So, for a single particle,

$$I = I_0 \frac{(1+\cos^2\theta)x^4 r^2}{2R^2}\left(\frac{\mu^2-1}{\mu^2+2}\right)^2 \tag{123}$$

Where, $x = 2\pi r/\lambda$, a parameter involving the size and wavelength of the particle and $r$ is the radius of the particle. Alongside the aforementioned, Rayleigh scattering requires $x \ll 1$. Averaging Eq. 123 over all angles gives us the Rayleigh scattering cross-section $\sigma_s$ [39].

For $x \gg 1$, the bodies are geometric shapes, scattering light as per their projected area.

For $x \cong 1$, We see the generalised phenomenon known as Mie Scattering.

## 9. ELECTROMAGNETIC RADIATION

By the 1800s, the link between thermodynamics, radiation, electromagnetism and optics was established.

It was seen that for all bodies interacting with radiation, the following relation was observed,

$$\alpha + \rho + \tau = 1 \tag{124}$$

$\alpha$, $\rho$ and $\tau$ correspond to absorptivity, reflectivity and transmissivity. For opaque bodies, $\tau = 0$.

In 1860, Gustav Kirchhoff showed that ratio of the radiating power ($E$) and absorptivity is a constant (intensity of radiation), independent to the nature of the body [55],

$$\frac{E(\lambda,T)}{\alpha} = E_b(\lambda,T) \tag{125}$$

To prove this law, Kirchhoff suggested that,
"The proof I am about to give of the law above stated, rests on the supposition that bodies can be imagined

---

[43] Newton [18] and Cavendish also had supposed something like this, but the observed deflection is half of that of the Newtonian prediction.

[44] Given as,

$$\alpha = \frac{E}{\mu}$$

Where, $\mu$ is the electrical dipole moment and $E$ is the electrical field strength.

which, for infinitely small thicknesses, completely absorb all incident rays, and neither reflect nor transmit any. I shall call such bodies perfectly black, or, more briefly, *black bodies*."

It was seen via Kirchhoff's law that for black bodies, $\alpha = \varepsilon = 1$, where $\varepsilon$ is emissivity. This made the black body a perfect absorber and emitter.

At thermal equilibrium, for a normal body with some index $i$, Eq. 125 showed,

$$\frac{E(\lambda, T, i)}{\alpha(\lambda, T, i)} = E_b(\lambda, T) = B_\lambda(T) \qquad (126)$$

Several attempts were made to describe the laws of black bodies and the $B_\lambda(T)$ function.

Josef Stefan showed in 1877, citing Tyndall's experiments on heating platinum wires, that the relation between the radiating power ($H$) and the fourth power of temperature [56] [57],

$$H \propto T_1^4 - T_2^4 \qquad (127)$$

Where $T_1$ and $T_2$ correspond to two temperature points.

Ludwig Boltzmann arrives at a similar law mathematically in 1884 [58],

$$H = \sigma \varepsilon A(T_1^4 - T_2^4) \qquad (128)$$

$\sigma$ is given as the Stefan-Boltzmann constant and $A$ is the area.

Wilhelm Wien took on the 'Kirchhoff problem' found a relation between the spectral energy densities $\Psi$ and temperature [59],

$$\frac{\Psi_1}{\Psi_2} = \left(\frac{\lambda_{P2}}{\lambda_{P1}}\right)^4 \qquad (129)$$

From here using the Stefan-Boltzmann Law (Eq. 128),

$$\frac{\lambda_{P2}}{\lambda_{P1}} = \frac{T_1}{T_2} = b \qquad (130)$$

The following gave a constant ($b$) from the peak wavelengths[45] and temperatures known as the Wien's constant.

He also provided a distribution law as follows [59] [60],

$$B(\lambda, T) = \frac{C_1}{\lambda^5} e^{-\frac{C_2}{\lambda T}} \qquad (131)$$

Here, $C_1$ and $C_2$ to two undetermined constants. This distribution law does hold for short wavelengths but fails as we move towards the longer wavelengths.

In 1900, Lord Rayleigh [61] [62] [63] and James Jeans [64] showed that,

$$I(\lambda, T) = \frac{2\pi c}{\lambda^4} k_B T \qquad (132)$$

Or in terms of spectral radiance[46],

$$B_\lambda(T) = \frac{2c}{\lambda^4} k_B T \qquad (133)$$

If we convert this in terms of frequency[47],

$$B_\nu(T) = \frac{2\nu^2}{c^2} k_B T \qquad (134)$$

As we take $\nu \to \infty$ or $\lambda \to 0$, $B_\nu(T)$ shoots to infinity in the UV frequencies. The consequence of this is that all objects should immediately emit all of their energies in the form of UV light (radiation). This came to be known as the *ultraviolet catastrophe*.

## 10. QUANTUM THEORY OF LIGHT

In the same year, Max Planck, a student of Kirchhoff, was working on perfecting the Wien Distribution (Eq. 130). He quotes that it is valid "due only to a restriction to short wave lengths and low temperatures" [65]

He considers the following assumption from vibrational energy ($U$)[48] and entropy ($S$) relations from both Eqs. 131 and 133,

$$\frac{d^2 S}{dU^2} = -\frac{constant}{U^2} \qquad (135)$$

This leads him to a revision of Eq. 130 as [65],

$$B(\lambda, T) = \frac{C_1}{\lambda^5} \frac{1}{e^{-C_2/\lambda T} - 1} \qquad (136)$$

However, Planck was unsatisfied with his assumption of Eq. 135. This is because if we do consider 48, we are taken back to Eq. 131 and the total entropy must depend on the energy of an individual resonator as well [66].

He shows that the system entropy $S_n$ is give by the following,

---

[45] The peak as per the blackbody radiation curve given by each blackbody corresponding to a temperature. See Fig. 35A in [69, p. 24].

[46] The relation to convert to spectral radiance is given as,

$$I(\lambda, T) = \pi B(\lambda, T)$$

[47] The relation is given as,

$$B_\lambda(T) = B_\nu(T) \frac{d\nu}{d\lambda}$$

Since, the units of $B_\lambda(T)$ and $B_\nu(T)$ are different and hence, Eq. 59 cannot be applied.

[48] Planck considered the work of a linear resonator to explain his refinement of Wien's law. For a n-numbered system, total entropy $S_n$ depends on total energy $U_n$ of the n linear resonators.

$$S_n = k \log W + C \qquad (137)$$

With $C$ and $k$ acting as constants. $W$ is the probability[49] so that n-number of resonators possess the energy $U_n$.

He also suggests that a continuous function cannot explain $U_n$ and hence a discrete quantity composed of an integral number of finite equal parts must be employed.

He shows,

$$U_n = P\epsilon \qquad (138)$$

$P$ represents some large number and $\epsilon$ is some undetermined energy element (quanta).

Using 'complexes', he turns Eq. 137 to,

$$S_n = k \log \frac{(n+P)^{n+P}}{n^n P^P} \qquad (139)$$

Simplifying and combining Eq. 138 [66],

$$S_n = k\left\{\left(1+\frac{U}{\epsilon}\right)\log\left(1+\frac{U}{\epsilon}\right) - \frac{U}{\epsilon}\log\frac{U}{\epsilon}\right\} \qquad (140)$$

For application of this to Wien's law, Planck considers the energy quanta $\epsilon$ to be related to frequency, and gives a relation,

$$\epsilon = h\nu \qquad (141)$$

$h$ is the famous Planck's constant. This radical assumption marked the beginning of quantum physics.

Based on this, he finds[50],

$$B(\nu, T) = \frac{2h\nu^3}{c^2}\frac{1}{e^{-hc/k_B T} - 1} \qquad (142)$$

And in terms of wavelength,

$$B(\lambda, T) = \frac{2hc^2}{\lambda^5}\frac{1}{e^{-hc/\lambda k_B T} - 1} \qquad (143)$$

This is known as *Planck's Distribution Law*.

In 1905, Einstein shows that in order to explain photoelectric phenomena[51], we must take aid of the Planckian quanta. He suggests that the kinetic energy of the electron leaving the solid will be [67],

$$h\nu - W_0 = KE_{electron} \qquad (144)$$

$W_0$ represents the work function[52] of the solid. Robert Millikan performed experiments on this which confirmed this theory.

Einstein suggested that many inaccuracies in the Maxwellian light models could be explained if energy of light waves where localised into quanta [68]. He showed the following relation between the wavelength and momentum,

$$p = \frac{h}{\lambda} \qquad (145)$$

A direct consequence based on Eq. 141 is that the energy of a light quanta is given as $E = pc$.

In 1923, Arthur Compton showed that an X-ray quantum with frequency $\nu$ colliding with an electron causes another quantum to emerge with a different frequency $\nu'$ [69]. This is given as [70],

$$\nu' = \frac{\nu}{1 + 2\frac{h\nu}{mc^2}\sin^2\frac{\theta}{2}} \qquad (146)$$

In terms of wavelength,

$$\lambda' = \lambda + 2\frac{h}{mc}\sin^2\frac{\theta}{2} \qquad (146)$$

Where, $h/mc$ is the *Compton Wavelength*.

In 1926, Gibert Lewis called these quanta, ***photons*** [71].

## 11. PHOTONS

*φῶς (phos),* the Greek for light, lent its name to the light particle. But this was not a conventional particle. It had both wave and particle features in it. Hence, the term 'light quantum particle' is used.

Photons ($\gamma$) are our best bet to explain all electromagnetic and optical phenomena till date. They are massless and chargeless bosons (as per theory) and have a spin of 1.

As per QFT, photons are the quantisation of the EM field and their spins correspond to the classical LH-RH polarisation systems.

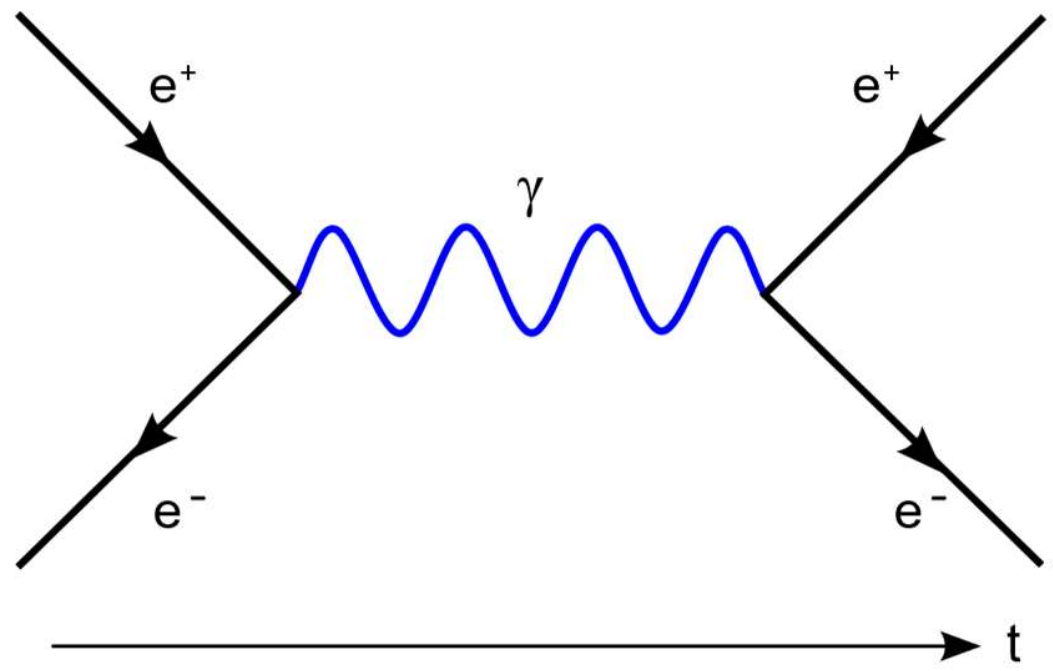

*Figure 7: Electron-Positron annihilation leads to formation of a virtual photon (blue). PC: Dirk Hünniger, Joel Holdsworth (from Wikipedia)*

[49] Planck suggests that electromagnetic theories do not possess such a probability and it is connected to the kinetic theory of gases.

[50] Exponential approximations for small wavelengths or long wavelengths give the Rayleigh-Jeans Law (Eqs. 133 and 134).

[51] See Gold leaf experiment.

[52] Given as $h\nu_0$. $\nu_0$ corresponds to the minimum EMR frequency required for electron excitation known as the threshold frequency.

Photons take part in many such particle interactions generating quasiparticles such as *polaritons*. This makes photons an indispensable part of the Standard Model of Particle Physics.

Photons are also used in lasers, electronics, with scopes in emerging fields like quantum computing.

## 12. CONCLUSION

In the end, we see the very humble and intuitive roots of light. The paper has tried to highlight notable advancements across history towards developing a robust and complete theory of it. Also seen is the hidden magnificence of the development of scientific understanding as more resources were available to the people with each formula and relation getting more and more intricate as we progress with time.

We started off with Indo-Hellenistic philosophies with minimal mathematics and complete intuition and ended with photons, which are dependent on some 2000+ years of intuition and rigorous mathematics.

The intricate beauty of such a ubiquitous phenomenon delights an enthusiast like me. If we are to go back to the opening lines and re-answer the question we asked,

"Light or electromagnetic radiation is not just what we see, but also what we can't see. It exists in all forms, with our eyes only being equipped to see a part of it. Light is a fundamental ingredient in this universe, for alongside being the carrier of the electromagnetic force, it is also the carrier of truth and clarity."

## 13. ACKNOWLEDGEMENTS

I am deeply thankful to my mentor Dr. Ivy Dutta[53], for guiding me and suggesting ideas on material. I am also grateful to Aniruddha Dutta and Anjishnu Dey for the detailed illustrations (Figures 1-6) that they have created for this endeavour.

A special mention goes to these books: [1] [8] [26] [39] [44] [49] [69] [72] for their general support in helping me understand parts of text by the great scientists.

## REFERENCES

[1] R. A. Crone, A History of Color: The Evolution of Theories of Light and Color, Springer Dordrecht, 1999.

[2] S. Singh, Fundamentals of Optical Engineering, New Delhi: Discovery Publishing House, 2009.

[3] R. Rashed, "A Pioneer in Anaclastics: Ibn Sahl on Burning Mirrors and Lenses," *Isis,* vol. 81, no. 3, pp. 464-491, 1990.

[4] Alhazen, Alhacen's Theory of Visual Perception, A. M. Smith, Ed., Philadelphia: American Philosophical Society, 2001.

[5] D. C. Lindberg, The Beginnings of Western Science, Chicago: The University of Chicago Press, 1992.

[6] H. Selin, Ed., Encyclopaedia of the History of Science, Technology, and Medicine in Non-Western Cultures, vol. 1, Springer, 2008, p. 1817.

[7] F. Cajori, A History of Physics in Its Elementary Branches, Macmillan, 1899.

[8] A. I. Sabra, Theories of Light from Descartes to Newton, 2nd ed., Cambridge University Press, 1981.

[9] J. K. McDonough, "Descartes' Optics," in *The Cambridge Descartes Lexicon*, L. Nolan, Ed., Cambridge University Press, 2015, pp. 197-199.

[10] R. Descartes, The World or Treatise on Light, 1633.

[11] S. Sakellariadis, "Descartes' Experimental Proof of the Infinite Velocity of Light and Huygens' Rejoinder," *Archive for History of Exact Sciences,* vol. 26, no. 1, pp. 1-12, 1982.

[12] R. J. MacKay and R. W. Oldford, "Scientific Method, Statistical Method and the Speed of Light," *Statistical Science,* vol. 15, no. 3, pp. 254-278, 2000.

[13] R. Foschi and M. Leone, "Galileo, measurement of the velocity of light, and the reaction times.," *Perception,* vol. 38, no. 8, pp. 1251-1259, 2009.

[14] J. H. Shea, "Ole Rømer, the speed of light, the apparent period of Io, the Doppler effect, and the dynamics of Earth and Jupiter," *Americal Journal of Physics,* vol. 66, no. 7, pp. 561-569, 1998.

[15] I. B. Cohen, "Roemer and the First Determination of the Velocity of Light (1676)," *Isis,* vol. 31, no. 2, pp. 327-379, 1940.

[16] F. Tuinstra, "Rømer and the Finite Speed of Light," *Physics Today,* vol. 57, no. 12, pp. 16-17, 2004.

[17] S. Fisher, The Stanford Encyclopedia of Philosophy, Spring ed., E. N. Zalta, Ed., Stanford: Metaphysics Research Lab, Stanford University, 2014.

[18] I. Newton, Opticks or, a Treatise of the Reflections, Refractions, Inflections, and Colours of Light, London, 1704.

[53] Department of Physics, Delhi Public School, Ruby Park, Kolkata

[19] K. M. Pedersen, “Leonhard Euler's Wave Theory of Light,” *Perspectives on Science,* vol. 16, no. 4, pp. 392-416, 2008.

[20] Z. Bechler, “Newton's Search for a Mechanistic Model of colour dispersion: A suggested interpretation,” *Archive for History of Exact Sciences,* vol. 11, no. 1, pp. 1-37, 1973.

[21] L. Foucault, “Thèse de physique. Sur les vitesses relatives de la lumière dans l’air et dans l’eau,” Paris, 1853.

[22] F. M. Grimaldi, Physicomathesis de lumine, coloribus et iride, 1665.

[23] R. Cocchini and G. Pelosi, “Diffraction: The First Recorded Observation,” *IEEE Antennas and Propagation Magazine,* vol. 32, no. 2, pp. 27-30, 1990.

[24] R. Hooke, Micrographia: or Some Physiological Descriptions of Minute Bodies Made by Magnifying Glasses. With Observations and Inquiries Thereupon., London: The Royal Society, 1665.

[25] C. Huygens, Traité de la Lumière: Où Sont Expliquées les Causes de ce qui Luy Arrive Dans la Reflexion & Dans la Refraction, Leiden: Van der Aa, 1690.

[26] NCERT, Physics - XII (Part - II), Unrationalised ed., New Delhi: NCERT, 2007.

[27] T. Young, “The Bakerian Lecture. On the theory of light and colours,” *Philosophical Transactions of the Royal Society of London,* vol. 92, pp. 12-48, 1802.

[28] T. Young, “The Bakerian Lecture. Experiments and Calculations relative to physical Optics.,” *Philosophical Transactions of the Royal Society of London,* vol. 94, pp. 1-16, 1804.

[29] D. Brewster, “On the laws which regulate the polarisation of light by reflexion from transparent bodies,” *Philosophical Transactions of the Royal Society of London,* vol. 105, pp. 125-159, 1815.

[30] C. Huygens, T. Young, F. Arago and A. J. Fresnel, The wave theory of light: memoirs of Huygens, Young and Fresnel., vol. 15, H. Crew, Ed., American Book Company, 1900.

[31] Z. J. Buchwald and C.-P. Yeang, “Kirchhoff’s theory for optical diffraction, its predecessor and subsequent development: the resilience of an inconsistent theory,” *Archive for History of Exact Sciences,* vol. 70, no. 5, pp. 463-511, 2016.

[32] J. R. Hofmann, André-Marie Ampère: Enlightenment and Electrodynamics, Cambridge University Press, 1996.

[33] M. Faraday, Faraday's Diary: Being the Various Philosophical Notes of Experimental Investigation, vol. IV, London: G. Bell and Sons Ltd., 1933.

[34] W. E. Weber and R. H. Kohlrausch, “Ueber die Elektricitätsmenge, welche bei galvanischen Strömen durch den Querschnitt der Kette fliesst,” *Annalen der Physik,* vol. 175, no. 9, pp. 10-25, 1856.

[35] W. E. Weber and R. H. Kohlrausch, Elektrodynamische Maassbestimmungen insbesondere Zurückführung der Stromintensitäts-Messungen auf mechanisches Maass, Springer Berlin Heidelberg, 1893.

[36] G. R. Kirchhoff, “Ueber die Bewegung der Elektricität in Leitern,” *Annalen der Physik,* vol. 178, no. 12, pp. 529-544, 1857.

[37] J. C. Maxwell, “On Physical Lines of Force,” *The London, Edinburgh, and Dublin Philosophical Magazine and Journal of Science,* vol. 23, no. 151, pp. 12-24, 1862.

[38] J. C. Maxwell, “A dynamical theory of the electromagnetic field,” *Philosophical transactions of the Royal Society of London,* vol. 155, pp. 459-512, 1865.

[39] R. P. Feynman, R. B. Leighton and M. Sands, The Feynman Lectures on Physics, 2nd ed., vol. 1, Dorling Kindersley India, 2008.

[40] J. Bradley, “Account of a new discovered motion of the fix'd stars,” *Philosophical Transactions of the Royal Society of London,* vol. 35, no. 406, pp. 637-661, 1728.

[41] K. F. Schaffner, Nineteenth-Century Aether Theories, Pergamon Press, 1972.

[42] G. G. Stokes, “On the aberration of light,” *The London, Edinburgh, and Dublin Philosophical Magazine and Journal of Science,* vol. 27, no. 177, pp. 9-15, 1845.

[43] A. A. Michelson and E. W. Morley, “On the Relative Motion of the Earth and the Luminiferous Ether,” *American Journal of Science,* vol. 34, no. 203, pp. 333-345, 1887.

[44] A. B. Gupta, Modern Atomic and Nuclear Physics, Books and Allied, 2011.

[45] H. A. Lorentz, “Electromagnetic phenomena in a system moving with any velocity less than that of light,” *Proceedings of the Royal Netherlands Academy of Arts and Sciences,* vol. 6, pp. 809-831, 1904.

[46] H. A. Lorentz, Attempt at a Theory of Electrical and Optical Phenomena in Moving Bodies, E. J. Brill, 1895.

[47] H. Poincaré, “On the Dynamics of the Electron,” *Rendiconti del Circolo matematico di Palermo,* vol. 21, p. 129–176, 1905-06.

[48] J. Reignier, “"The birth of special relativity." One more essay on the subject,” 2000. [Online]. Available: https://arxiv.org/pdf/physics/0008229.pdf. [Accessed 2023 July 25].

[49] A. Einstein, Relativity, 2019 ed., Wilco Publishing House, 1916.

[50] A. Einstein, “On the Electrodynamics of Moving Bodies,” *Annalen der Physik,* vol. 17, pp. 891-921, 1905.

[51] A. Einstein, H. Minkowski, S. N. Bose and M. N. Saha, The principle of relativity; original papers, Kolkata: Calcutta University Press, 1920.

[52] A. S. Eddington, F. W. Dyson and C. R. Davidson, “Philosophical Transactions of the Royal Society of London. Series A,” *A determination of the deflection of light by the sun's gravitational field, from observations made at the total eclipse of May 29, 1919,* vol. 220, no. 571-581, pp. 291 - 333, 1920.

[53] J. Tyndall, “On the blue colour of the sky, the polarization of skylight, and on the polarization of light by cloudy matter generally,” *Proceedings of the Royal Society of London,* vol. 17, pp. 223-233, 1869.

[54] J. W. Strutt, “On the Transmission of Ligld through an Atmosphere containing Small Particles in Suspenslon, and on the Origin of the Blue of the Sky.,” *The London, Edinburgh, and Dublin Philosophical Magazine and Journal of Science,* vol. 47, no. 287, pp. 375-384, 1899.

[55] G. R. Kirchhoff, “On the Relation between the Radiating and Absorbing Powers of different Bodies for Light and Heat,” *The London, Edinburgh, and Dublin Philosophical Magazine and Journal of Science,* vol. 20, no. 130, pp. 1-21, 1860.

[56] J. Stefan, “Über die Beziehung zwischen der Wärmestrahlung und der Temperatur,” *Sitzungsberichte der Kaiserlichen Akademie der Wissenschaften: Mathematisch-Naturwissenschaftliche Classe ,* vol. 79, pp. 391-428, 1879.

[57] J. Satterly, “Stefan's Radiation Law,” *Nature,* vol. 157, no. 3996, p. 737, 1946.

[58] L. E. Boltzmann, “Ableitung des Stefan'schen Gesetzes, betreffend die Abhängigkeit der Wärmestrahlung von der Temperatur aus der electromagnetischen Lichttheorie,” *Annalen der Physik und Chemie,* vol. 258, no. 6, pp. 291-294, 1884.

[59] J. Crepeau, “A brief history of the $T^4$ radiation law,” *Heat Transfer Summer Conference,* vol. 43567, pp. 59-65, 2009.

[60] W. Wien, “Ueber die Energievertheilung im Emissionsspectrum eines schwarzen Körpers,” *Annalen der Physik,* vol. 294, no. 8, pp. 662-669, 1896.

[61] J. W. Strutt, “Remarks upon the Law of Complete Radiation,” in *Scientifc Papers*, vol. 4, Cambridge University Press, 1900, pp. 483-485.

[62] J. W. Strutt, “The Dynamical Theory of Gases and of Radiation,” *Nature,* vol. 72, no. 1855, pp. 54-55, 1905.

[63] J. W. Strutt, “The Constant of Radiation as Calculated from Molecular Data,” *Nature,* vol. 72, no. 1863, pp. 243-244, 1905.

[64] J. H. Jeans, “The Dynamical Theory of Gases,” *Nature,* vol. 71, no. 1852, p. 607, 1905.

[65] M. Planck, “On an Improvement of Wien’s Equation for the Spectrum,” *Transactions of the German Physical Society,* vol. 2, pp. 202-204, 1900.

[66] M. Planck, “Ueber das Gesetz der Energieverteilung im Normalspectrum,” *Annalen der Physik,* vol. 4, no. 3, pp. 553-563, 1901.

[67] A. Einstein, “ Über einen die Erzeugung und Verwandlung des Lichtes betreffenden heuristischen Gesichtspunkt,” *Annalen der Physik,* vol. 17, no. 6, p. 132–148, 1905.

[68] A. Einstein, “The Development of Our Views on the Composition and Essence of Radiation,” *Physikalische Zeitschrift,* vol. 10, no. 22, pp. 817-825, 1909.

[69] E. H. Wichman, Berkeley Physics Course Volume 4: Quantum Physics (In SI Units), McGraw Hill, 2011.

[70] A. H. Compton, “A quantum theory of the scattering of X-rays by light elements,” *Physical Review,* vol. 21, no. 5, pp. 483-502, 1923.

[71] G. N. Lewis, “The Conservation of Photons,” *Nature,* vol. 118, no. 2981, pp. 874-875, 1926.

[72] B. D. Gupta, Mathematical Physics, Vikas Publishing House, 1980.